\documentclass[sigconf]{acmart}

\copyrightyear{2026}
\acmYear{2026}
\setcopyright{cc}
\setcctype{by}
\acmConference[WWW '26]{Proceedings of the ACM Web Conference 2026}{April 13--17, 2026}{Dubai, United Arab Emirates}
\acmBooktitle{Proceedings of the ACM Web Conference 2026 (WWW '26), April 13--17, 2026, Dubai, United Arab Emirates}
\acmPrice{}
\acmDOI{10.1145/3774904.3792075}
\acmISBN{979-8-4007-2307-0/2026/04}

\usepackage{xcolor}
\usepackage{booktabs}
\usepackage{bbding}
\usepackage{xcolor}
\newcommand{\cmark}{{\color{green}\CheckmarkBold}}
\newcommand{\xmark}{{\color{red}\XSolidBrush}}
\usepackage{amsthm}

\usepackage[ruled,vlined]{algorithm2e}
\usepackage{multirow}
\usepackage[font=small]{caption}
\usepackage{balance}
\usepackage{adjustbox}
\usepackage{mathtools}
\usepackage{parskip}  

\theoremstyle{definition}

\theoremstyle{remark}

\usepackage{enumitem}
\usepackage{pifont}
\usepackage{threeparttable}
\usepackage{xcolor,colortbl}
\definecolor{blue}{HTML}{3331D7}
\definecolor{orange}{HTML}{E57932}
\definecolor{purple}{HTML}{663399}
\definecolor{green}{HTML}{41805E}
\definecolor{rose}{HTML}{C71585}
\definecolor{crimson}{HTML}{DC143C}
\definecolor{grey}{HTML}{505050}
\definecolor{skyblue}{RGB}{203, 221, 245}
\begin{document}
\renewcommand{\arraystretch}{0.9}

\title[What Should I Cite? A RAG Benchmark for Academic Citation Prediction]{What Should I Cite? A RAG Benchmark for\\ Academic Citation Prediction}

\author{Leqi Zheng}
\authornote{These authors contributed equally to this research.}
\orcid{0009-0001-8012-100X}
\affiliation{%
  \institution{Tsinghua University}
  \city{Beijing}
  \country{China}}
\email{zhenglq24@mails.tsinghua.edu.cn}
  
\author{Jiajun Zhang}
\authornotemark[1]
\orcid{0009-0005-1379-2005}
\affiliation{%
  \institution{University of Science and Technology of China}
  \city{Hefei}
  \country{China}}
\email{zhangjiajun519@mail.ustc.edu.cn}

\author{Canzhi Chen}
\authornotemark[1]
\orcid{0009-0001-3888-5719}
\email{chencanzhi@bit.edu.cn}
\affiliation{%
  \institution{Beijing Institute of Technology}
  \city{Beijing}
  \country{China}}

\author{Chaokun Wang}
\orcid{0000-0002-2986-2574}
\affiliation{%
  \institution{Tsinghua University}
  \city{Beijing}
  \country{China}}
\authornote{Chaokun Wang is the corresponding author.}
\email{chaokun@tsinghua.edu.cn}

\author{Hongwei Li}
\orcid{0009-0008-8813-2433}
\affiliation{%
  \institution{Tsinghua University}
  \city{Beijing}
  \country{China}}

\author{Yuying Li}
\orcid{0009-0007-0817-4004}
\affiliation{%
  \institution{Tsinghua University}
  \city{Beijing}
  \country{China}}
  
\author{Yaoxin Mao}
\orcid{0009-0008-1386-7662}
\affiliation{%
  \institution{Beijing Institute of Technology}
  \city{Beijing}
  \country{China}}

\author{Shannan Yan}
\orcid{0009-0008-2278-9977}
\affiliation{%
  \institution{Tsinghua University}
  \city{Beijing}
  \country{China}}

\author{Zixin Song}
\orcid{0009-0008-9111-3866}
\affiliation{%
  \institution{Tsinghua University}
  \city{Beijing}
  \country{China}}

\author{Zhiyuan Feng}
\orcid{0009-0002-4610-422X}
\affiliation{%
\institution{Tsinghua University}
\city{Beijing}
\country{China}}

\author{Zhaolu Kang}
\orcid{0009-0000-1163-1615}
\affiliation{%
  \institution{Peking University}
  \city{Beijing}
  \country{China}}

\author{Zirong Chen}
\orcid{0009-0006-6907-8289}
\affiliation{%
  \institution{Tsinghua University}
  \city{Beijing}
  \country{China}}

\author{Hang Zhang}
\orcid{0009-0009-5918-3183}
\affiliation{%
  \institution{Tsinghua University}
  \city{Beijing}
  \country{China}}

\author{Qiang Liu}
\orcid{0000-0002-9233-3827}
\affiliation{%
  \institution{Institute of Automation, Chinese Academy of Sciences}
  \city{Beijing}
  \country{China}}

\author{Liang Wang}
\orcid{0000-0001-5224-8647}
\affiliation{%
  \institution{Institute of Automation, Chinese Academy of Sciences}
  \city{Beijing}
  \country{China}}

\author{Ziyang Liu}
\orcid{0009-0007-4238-1533}
\affiliation{%
  \institution{Tsinghua University}
  \city{Beijing}
  \country{China}}

\renewcommand{\shortauthors}{Leqi Zheng et al.}

\begin{abstract}
With the rapid growth of Web-based academic publications, more and more papers are being published annually, making it increasingly difficult to find relevant prior work. Citation prediction aims to automatically suggest appropriate references, helping scholars navigate the expanding scientific literature. Here we present \textbf{CiteRAG}, the first comprehensive retrieval-augmented generation (RAG)-integrated benchmark for evaluating large language models on academic citation prediction, featuring a multi-level retrieval strategy, specialized retrievers, and generators. Our benchmark makes four core contributions: (1) We establish two instances of the citation prediction task with different granularity. Task 1 focuses on coarse-grained list-specific citation prediction, while Task 2 targets fine-grained position-specific citation prediction. To enhance these two tasks, we build a dataset containing 7,267 instances for Task 1 and 8,541 instances for Task 2, enabling comprehensive evaluation of both retrieval and generation. (2) We construct a three-level large-scale corpus with 554k papers spanning many major subfields, using an incremental pipeline. (3) We propose a multi-level hybrid RAG approach to citation prediction, fine-tuning embedding models with contrastive learning to capture complex citation relationships, paired with specialized generation models. (4) We conduct extensive experiments across state-of-the-art language models, including closed-source APIs, open-source models, and our fine-tuned generators, demonstrating the effectiveness of our framework. Our open-source toolkit enables reproducible evaluation and focuses on academic literature, providing the first comprehensive evaluation framework for citation prediction and serving as a methodological template for other scientific domains. Our source code and data are released at \url{https://github.com/LQgdwind/CiteRAG}. 

\end{abstract}



\maketitle

\section{Introduction}

With the rapid advancement of scientific research on the Web, an unprecedented number of academic papers are published every day, forming a massive and ever-growing body of online scholarly knowledge. At the heart of this Web-based scientific corpus lies the citation network, which connects individual studies through citation relationships and continuously integrates new findings into the existing body of science. As citation networks play a central role in the evolution of scientific knowledge, citation prediction, which aims to forecast the prior studies a new paper will cite, has attracted growing research attention. From an individual researcher’s perspective, citation prediction can facilitate the discovery of relevant prior work, improve the quality of scholarly references, and foster interdisciplinary innovation. From a broader perspective, it provides a powerful analytical tool for computational social science \cite{Lazer2009CSS, Hofman2021Integrating, Ciotti2015Homophily}, revealing the hidden structure of citation networks and offering valuable insights into how knowledge propagates, transforms, and gives rise to innovation.

Despite extensive research on citation prediction~\cite{cohan2019structural, zhang2023changes}, there remains a notable absence of a comprehensive benchmark for evaluating models in this domain. Existing benchmarks suffer from several key limitations: (1) \textbf{oversimplified task formulations}, as real-world users often query related studies using specific sections of a paper, such as the introduction, methodology, or experimental details, to identify the most relevant and fine-grained prior work~\cite{cohan2019structural}, whereas current benchmarks typically treat citation prediction as a binary classification or document-level retrieval problem within a fixed corpus, an oversimplification that diverges from practical scenarios; (2) \textbf{ignorance of the hierarchical structure of scientific literature}, since scientific texts are inherently layered and rhetorically organized~\cite{teufel1999annotation,zhang2023changes,kashyap2023scientific}, yet current benchmarks often flatten this complexity into a textual knowledge graph assuming full-text availability, thus neglecting the multi-level organization of scientific discourse; (3) \textbf{suboptimal methodological design}, as most existing approaches rely on fine-tuning simple embedding-based retrievers~\cite{reimers2019sentence,gao2021simcse,lee2019latent}, which struggle to capture the nuanced, structural nature of citation relationships and tend to lose retrieval effectiveness as the candidate set expands~\cite{lee2019latent}; (4) \textbf{a lack of systematic evaluation framework}, as current benchmarks typically focus on citation network construction and simple embedding-based methods while failing to provide a comprehensive evaluation of more advanced approaches such as retrieval-augmented generation (RAG)~\cite{lewis2020retrieval,izacard2021leveraging}. Other platforms, such as Deep Research-style agents \cite{zheng2025deepresearcher}, focus on literature exploration rather than serving as standardized benchmarks, and cannot serve as formal benchmarks for citation prediction.

\begin{table}[t]
\centering
\small
\caption{Comparison of our proposed benchmark with existing citation datasets \cite{Cohan2020SPECTER, Saier2023unarXive, Tilwani2024REASONS, ajith2024litsearch} across five key dimensions. Multidimensional Evaluation means our work assesses not just retrieval or classification, but also hallucination and diversity in RAG systems.}
\setlength{\tabcolsep}{3pt}
\resizebox{\linewidth}{!}{
\begin{tabular}{lccccc}
\toprule
\textbf{Dataset} & \textbf{SciDocs} & \textbf{unarXive} & \textbf{REASONS} &\textbf{LitSearch} & \textbf{Ours} \\
\midrule
\textbf{Multi-domain} & \cmark & \cmark & \cmark & \cmark& \cmark \\
\textbf{Structured Full-Text} & \xmark & \cmark & \xmark & \xmark & \cmark \\
\textbf{Support for RAG Tasks} & \xmark & \xmark & \cmark & \cmark & \cmark \\
\textbf{Multi-level Corpus} & \xmark & \xmark & \xmark & \xmark & \cmark \\
\textbf{Multidimensional Evaluation} & \xmark & \xmark & \xmark & \xmark & \cmark \\
\bottomrule
\end{tabular}
}
\vspace{-4mm}
\label{tab:benchmark_comparison}
\end{table}

To address this gap, we propose a comprehensive benchmark for citation prediction that encompasses two complementary tasks operating at different levels of granularity. List-specific citation prediction (Task 1) evaluates a system’s ability to generate complete reference lists for papers. Position-specific citation prediction (Task 2) assesses a system’s capability to predict precise citations for individual reference placeholders within the paper text. Building on these two tasks, we collected a large corpus of scientific texts from Google Scholar, resulting in a multidisciplinary, hierarchical corpus of 554k papers. Each paper is represented with multi-level textual information and associated with a large candidate set. To facilitate the evaluation of citation prediction tasks, we further curated 7,267 samples for Task 1 and 8,541 samples for Task 2. 

Given the massive size of the candidate pool and the limitations of conventional embedding-based methods in capturing citation relations, we design a multi-level hybrid retrieval-augmented generation (RAG) approach to citation prediction. Our approach integrates a carefully fine-tuned, state-of-the-art 8B retriever with a suite of task-specific large language models (LLMs). To establish a solid foundation for future research, we conduct a comprehensive evaluation of our RAG pipeline alongside a wide range of baseline methods, including both open- and closed-source LLMs as well as traditional embedding-based retrievers. The results highlight the intrinsic complexity of the citation prediction tasks and confirm the effectiveness of our proposed approach. Finally, we release an open-source toolkit that enables reproducible evaluation and provides the first comprehensive framework dedicated to citation prediction. Table~\ref{tab:benchmark_comparison} summarizes the key features of our benchmark compared to existing datasets. Our benchmark is multidisciplinary and supports RAG-based baselines. It also includes the evaluation for hallucination and diversity, is structured as a hierarchical RAG pipeline, and provides a reproducible open-source toolkit.

In summary, our contributions are fourfold:

(1) \textbf{Task Definition and Granularity Distinction.} We define two novel instances of the citation prediction tasks: Task 1, focusing on coarse-grained list-specific prediction, and Task 2, targeting fine-grained position-specific prediction. The dataset we built contains 7,267 instances for Task 1 and 8,541 instances for Task 2, enabling comprehensive evaluation of both retrieval and generation across different task granularities (Section~\ref{sec:Task_def} and Section~\ref{sec:corpus}).

(2) \textbf{Multi-Level Incremental Corpus Construction.} We construct a large-scale hierarchical corpus with 554,719 papers, built using an incremental pipeline that supports continuous updates. This enables flexible access to multi-level information and scalable expansion of the academic knowledge base for citation prediction tasks (Section~\ref{sec:corpus}).

(3) \textbf{Multi-Level Hybrid Retrieval and Generation Framework.} We propose a multi-level hybrid RAG approach, combining specialized retrievers tuned for different corpus levels with generation models fine-tuned to capture complex citation relationships. This multi-level hybrid RAG approach improves both retrieval precision and context-aware citation generation compared to single-level methods (Section~\ref{sec:retriever}).

(4) \textbf{Open-Source RAG Evaluation Toolkit.} We establish a comprehensive evaluation framework covering the full citation prediction pipeline, from retrieval effectiveness to end-to-end generation performance. The toolkit provides standardized metrics, evaluation protocols, and extensive comparison across state-of-the-art models, including closed-source APIs, open-source LLMs, and our fine-tuned generators, enabling reproducible benchmarking and facilitating further research (Section~\ref{sec:eval}, Section~\ref{sec:experimental_setup} and Section~\ref{sec:results}).

\section{Preliminaries and Task Definition}
\label{sec:Task_def}
We formulate academic citation prediction as a retrieval-augmented generation (RAG) problem operating at two distinct granularities to evaluate the capabilities of different large language models. The first task addresses coarse-grained citation prediction for entire papers, while the second focuses on fine-grained citation prediction for specific textual positions.

\subsection{Task 1: List-Specific Citation Prediction}

The list-specific citation prediction task aims to predict the complete reference list for a given academic paper. Let $\mathcal{C} = \{c_1, c_2, ..., c_n\}$ denote the corpus of available papers. Given a query paper $q$ with title $T_q$ and abstract $A_q$, the task generates a ranked list of reference papers:

\begin{equation}
R_q = (r_1, r_2, ..., r_k), \quad r_i \in \mathcal{C}
\end{equation}

where $k$ is the number of predicted citations and the ranking function $f: (T_q, A_q) \rightarrow R_q$ maps the input to an ordered sequence reflecting citation relevance.

This coarse-grained formulation employs a fixed-length prefix chunking strategy, using title and abstract (typically 200-400 tokens) to represent each paper. This design addresses practical constraints in academic literature where full-text access is frequently restricted by paywalls or copyright, while titles and abstracts remain universally accessible through bibliographic databases. The approach balances computational efficiency with real-world applicability across heterogeneous corpus conditions.

\subsection{Task 2: Position-Specific Citation Prediction}

The position-specific citation prediction task operates at finer granularity, predicting the specific paper to cite at individual reference placeholders within the text. Let $q$ be a query paper with title $T_q$, abstract $A_q$, and textual sections $S_q$ containing reference placeholders $\{[ref]_1, [ref]_2, ..., [ref]_m\}$. For each reference placeholder $[ref]_i$ with surrounding context $\mathcal{X}_i \subset S_q$, the task predicts the appropriate citation:

\begin{equation}
\hat{r}_i = g(T_q, A_q, \mathcal{X}_i; \mathcal{C}), \quad \hat{r}_i \in \mathcal{C}
\end{equation}

where $g$ is the prediction function that selects the most relevant paper from corpus $\mathcal{C}$ based on the local context $\mathcal{X}_i$ and paper metadata.

This fine-grained formulation requires understanding the immediate textual context surrounding each reference placeholder. The prediction system must analyze specific claims, methodologies, or concepts being discussed, identify the required citation support type, and select the most appropriate reference. This prediction system reflects the actual writing process where authors make precise decisions about citations at each point in their argumentation, considering chronological precedence, methodological relevance, and argumentative support.

\begin{figure*}[t]
    \centering
    \includegraphics[width=1.0\linewidth]{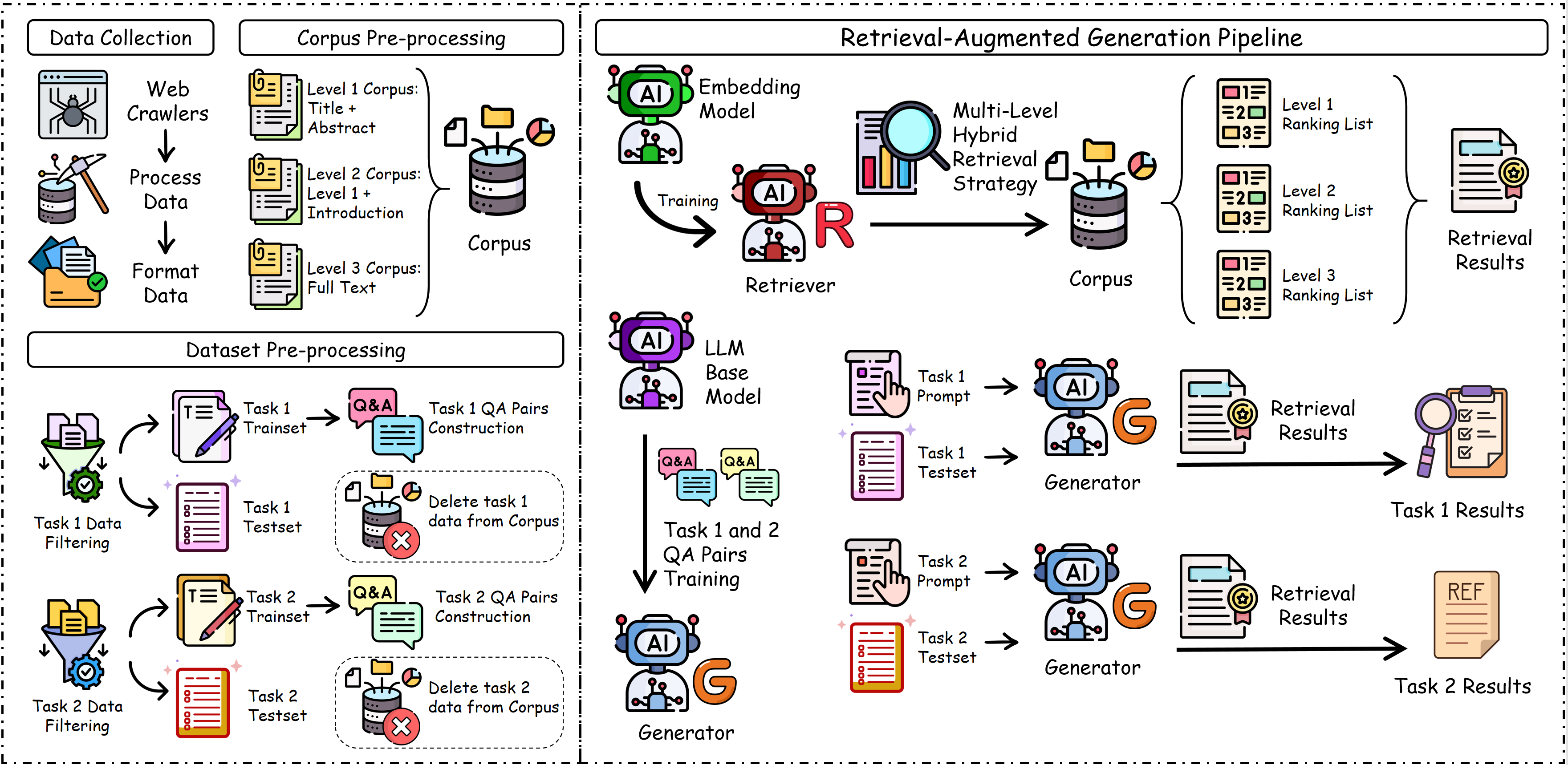}
\vspace{-3mm}
\caption{Overview of the CiteRAG benchmark pipeline comprising three stages: (1) Data Collection and Corpus Pre-processing constructs a three-level hierarchical corpus from web-crawled papers; (2) Dataset Pre-processing filters and formats data into Task 1 and Task 2 QA pairs with test instances removed from the corpus; (3) Retrieval-Augmented Generation Pipeline trains embedding models and retrievers, then applies multi-level hybrid retrieval to feed generators for task-specific citation prediction and evaluation.}
    \vspace{-3mm}
    \label{fig:pipeline}
\end{figure*}

\section{Corpus and Dataset Construction}
\label{sec:corpus}

This section details the construction of the RAG benchmark dataset, which encompasses both the core citation prediction dataset and comprehensive external knowledge corpora. The dataset construction process involves three primary phases: systematic data collection, rigorous preprocessing and quality control measures, and the creation of specialized retrieval corpora to support end-to-end citation prediction tasks. 

\subsection{Data Collection}

To construct a comprehensive corpus for academic citation prediction, we systematically collected computer science papers from Google Scholar over the past decade, ensuring broad temporal coverage of evolving research trends and citation patterns.  

The collection framework implements dynamic query strategies that adapt to the volume characteristics of different research areas. For smaller subcategories containing fewer than 2,000 papers, direct queries with year-based filtering provide efficient retrieval. Medium-sized categories with 2,000 to 12,000 papers employ month-by-month querying. Large categories exceeding 12,000 papers utilize week-by-week collection strategies to prevent data loss due to API limitations. This adaptive approach successfully collected papers across 10 computer science subcategories, including AI \& ML, DB \& DM, Systems, and others.

\subsection{Corpus and Dataset Pre-processing}
\label{sec:preprocessing}
\subsubsection{Corpus Pre-processing}

The corpus preprocessing pipeline implements a three-tier hierarchical structure for multi-granular citation prediction evaluation. Level 1 contains domain classifications, titles, and abstracts for rapid relevance assessment. Level 2 adds introduction sections to Level 1 components, enabling a deeper understanding of research motivation and methodology. Level 3 incorporates full-text content with references removed, plus conclusion sections, providing comprehensive coverage while maintaining clean input for generation models. The preprocessing framework employs pattern-matching algorithms to extract introduction sections and removes citation markers to eliminate prediction interference. All levels maintain consistent metadata, including authorship, publication year, and venue information, supporting both coarse-grained and fine-grained citation prediction tasks through appropriate information density matching.

\subsubsection{Dataset Pre-processing}

The List-Specific Citation Prediction (Task 1) preprocessing establishes quality control measures to identify papers with sufficient citation information for meaningful evaluation. We filter out papers with non-standard citations (including duplicate citations, citation errors, etc.) and retain essential components, including identifiers, titles, abstracts, and complete reference collections, while removing extraneous metadata. This approach ensures adequate citation diversity for comprehensive list-level prediction assessment.

The Position-Specific Citation Prediction (Task 2) preprocessing employs a more sophisticated multi-stage approach to create fine-grained citation scenarios. We also initially filter out papers with non-standard citations, and then analyze citation distribution patterns to identify the three most frequently cited sections within each paper. For these high-citation sections, we preserve the top three most frequent citations with valid identifiers while removing other citation markers, creating clean textual contexts with precisely positioned reference placeholders. Papers where any single reference appears more than ten times within a section are excluded to prevent citation bias.

To ensure rigorous evaluation integrity and prevent data contamination, all papers successfully processed and incorporated into either Task 1 or Task 2 datasets are permanently removed from the original corpus.

\begin{figure*}[!t]
    \centering
    \includegraphics[width=0.95\linewidth]
    {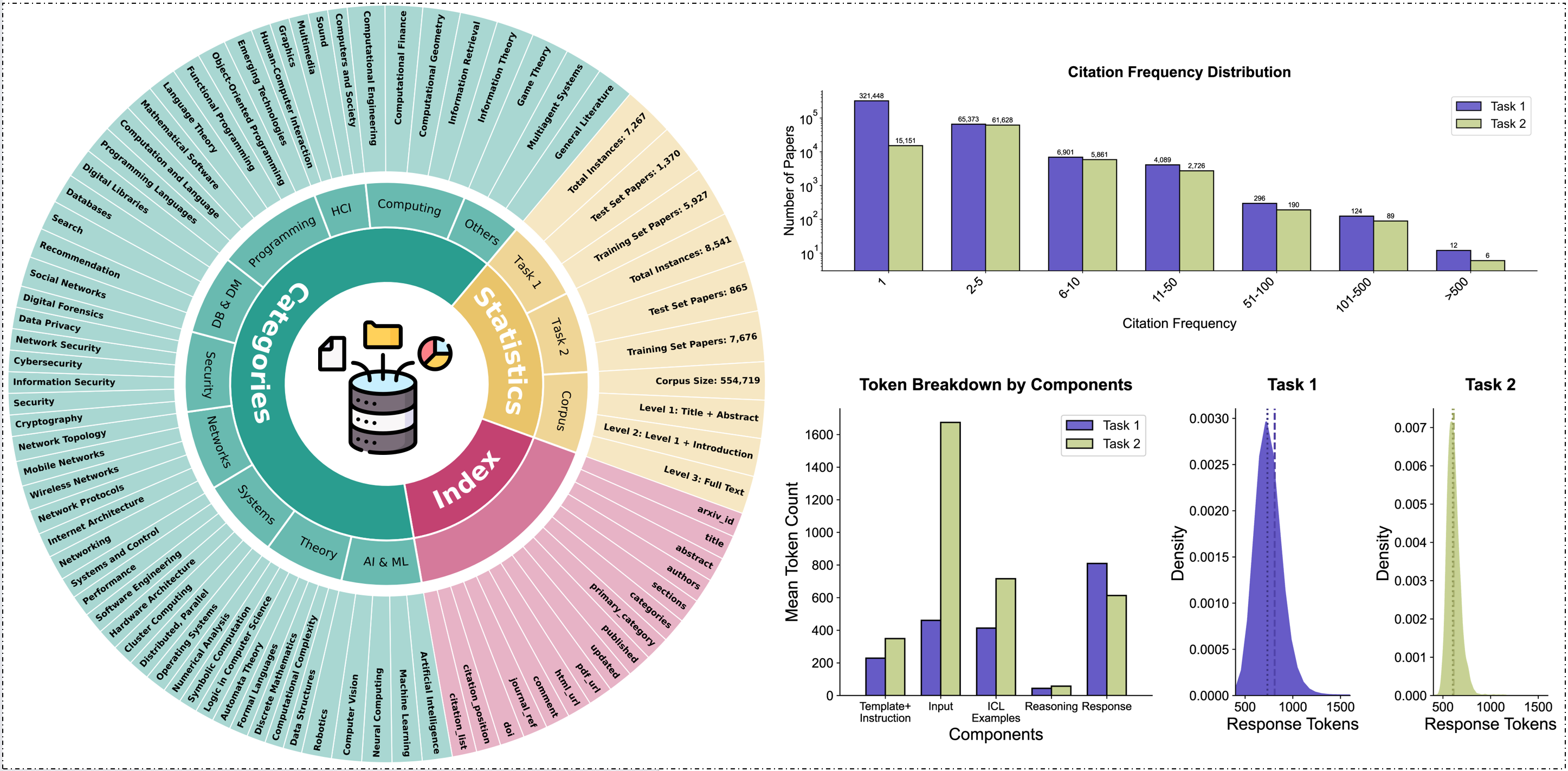}
    \vspace{-3mm}
\caption{Corpus and dataset statistics. The left sunburst chart shows three-level organization with 554,719 total papers across 10 fields, training/test distributions, and metadata structure. The upper right panels display citation frequency distributions for Task 1 and Task 2. The lower right panels show token composition breakdowns and response token length distributions for both tasks.}
    \vspace{-3mm}
    \label{fig:corpus}
\end{figure*}

\subsection{Corpus and Dataset Statistics}

The comprehensive statistics of our benchmark corpus and evaluation datasets are presented in Figure~\ref{fig:corpus}. 
The datasets are formatted as question-answer pairs following the ChatML convention detailed in Appendix~\ref{appendix:QA}, enabling seamless integration with standard language model training frameworks.

\section{Retriever Design}
\label{sec:retriever}

Academic papers exhibit hierarchical textual structures with varying information density across different granularity levels. To address the challenges of multi-granular information access and limited full-text availability, we develop a specialized retriever optimized for academic citation prediction tasks.

\subsection{Multi-Level Hybrid Retrieval Strategy}

We implement a retriever system $\mathcal{M}$ that directly leverages the three-level corpus hierarchy constructed in Section~\ref{sec:corpus}. The retriever system is formally defined as $\mathcal{M} = (\text{Encode}, \text{Retrieve}, S)$ where $\text{Encode}$ represents the embedding function, $\text{Retrieve}$ denotes the similarity search function, and $S$ is the ranking fusion strategy. This architecture enables multi-granular information access across the pre-existing corpus levels $\mathcal{D}^{1}, \mathcal{D}^{2}, \mathcal{D}^{3}$, which correspond to Level 1, Level 2, and Level 3 corpus representations respectively.

Our chunking strategy employs title and abstract content as standardized query input, maintaining consistency with Task 1 requirements while enabling efficient retrieval across all corpus granularities. Given query paper $q$ consisting of title and abstract, the retrieval system performs parallel top-$k$ similarity search across the three pre-established corpus levels:

\begin{align}
\mathcal{R}^{1}_k &= \text{Retrieve}(q, \mathcal{D}^{1}), \\
\mathcal{R}^{2}_k &= \text{Retrieve}(q, \mathcal{D}^{2}), \\
\mathcal{R}^{3}_k &= \text{Retrieve}(q, \mathcal{D}^{3}).
\end{align}

Results are merged using reciprocal rank fusion:
\begin{align}
\mathcal{R}^*_k = S\big(\mathcal{R}^{1}_k, \mathcal{R}^{2}_k, \mathcal{R}^{3}_k\big).
\end{align}

This architecture enhances retrieval diversity while maintaining robustness across varying corpus availability conditions.

\subsection{Contrastive Learning Fine-tuning}

Standard embedding models are inadequately optimized for academic citation relationships, which encompass complex interdependencies beyond semantic similarity. Then, we fine-tune Qwen-3-embedding-8B as CitationRetriever-8B using contrastive learning to capture citation-specific patterns.

From the training corpus, we select 30,000 query papers and their cited references, constructing 400,000 query-positive pairs $(q, d^+)$ across all granularity levels. We train the model using InfoNCE loss with temperature scaling:

\begin{align}
\mathcal{L} = -\log \frac{\exp(\text{sim}(q, d^+)/\tau)}{\sum_{d^- \in \mathcal{N}(q)} \exp(\text{sim}(q, d^-)/\tau) + \exp(\text{sim}(q, d^+)/\tau)}
\end{align}

where $\text{sim}(\cdot, \cdot)$ denotes cosine similarity, $\tau$ is the temperature parameter, and $\mathcal{N}(q)$ represents negative samples from approximate nearest neighbors and in-batch negatives.

The resulting retriever model, CitationRetriever-8B, demonstrates enhanced capability in identifying complex citation relationships beyond surface-level semantic matching.

\section{Evaluation Metrics}
\label{sec:eval}
The evaluation framework employs distinct metrics tailored to each component of the RAG pipeline, ensuring comprehensive assessment of both retrieval effectiveness and generation quality. The metrics are designed to capture different aspects of system performance while maintaining consistency with established practices in information retrieval and citation analysis.

\subsection{Retriever Evaluation Metrics}

We employ two standard information retrieval metrics to evaluate retriever performance: Recall@k and Mean Reciprocal Rank@k (MRR@k). The detailed mathematical formulations are provided in Appendix~\ref{appendix:standard_metrics}.

\subsection{Task 1 Evaluation Metrics}

Task 1 evaluation employs three standard metrics for reference generation: Recall@k, Normalized Discounted Cumulative Gain@k (NDCG@k), and Hit@k. The complete definitions are provided in Appendix~\ref{appendix:standard_metrics}.

\subsection{Task 2 Evaluation Metrics}

Task 2 evaluation employs Position-Aware Citation Accuracy@k (PACA@k), a novel metric we propose to assess position-specific citation prediction quality.

\textbf{Position-Aware Citation Accuracy@k (PACA@k)} accounts for both correctness and ranking positions:
\begin{equation}
\text{PACA@k} = \frac{1}{N} \sum_{i=1}^{N} \sum_{j \in \text{correct@k}} \left(1 - \frac{\text{rank}_j - 1}{k}\right)
\end{equation}
where $N$ is the total number of reference placeholders, correct@k denotes correct predictions within top-$k$, and $\text{rank}_j$ is the position of correct prediction $j$.

PACA@k assigns higher scores to correct predictions at earlier ranks, reflecting that users typically consider only top-ranked suggestions when selecting citations for specific positions.

\subsection{Diversity and Quality Metrics}

Beyond retrieval accuracy, we propose two complementary metrics to assess the quality and diversity of predicted reference lists: Citation Diversity Entropy and Hallucination Rate.

\textbf{Citation Diversity Entropy (CDE)} measures the diversity of predicted citations across different research categories. For list-specific citation prediction (Task 1), given a predicted reference list with citations distributed across $C$ categories, let $p_i$ denote the proportion of citations in category $i$. The Citation Diversity Entropy is computed as:

\begin{equation}
\text{CDE} = -\sum_{i=1}^{C} p_i \log_2(p_i)
\end{equation}

where higher entropy values indicate more diverse citation patterns across research areas, while lower values suggest concentration in fewer categories. The maximum possible entropy is $\log_2(C)$, achieved when citations are uniformly distributed across all categories.

\textbf{Hallucination Rate (Halluc.)} quantifies the proportion of predicted citations that refer to non-existent papers, indicating model reliability. For a predicted reference list $P$, let $V$ denote the set of papers that can be verified to exist in the real world through bibliographic databases. The Hallucination Rate is defined as:

\begin{equation}
\text{Halluc.} = \frac{|P \setminus V|}{|P|} \times 100\%
\end{equation}

where $P \setminus V$ represents predictions that cannot be verified as real papers. Lower hallucination rates indicate better model grounding and reduced generation of fabricated references, which is critical for maintaining academic integrity in citation prediction systems.

\section{Experimental Setup}
\label{sec:experimental_setup}

\begin{table*}[htbp]
\caption{Performance comparison across citation prediction tasks. \textbf{Bold} indicates best overall performance, \underline{underline} indicates best baseline performance. * indicates statistical significance ($p < 0.05$).}
\vspace{-2mm}
\centering
\begin{adjustbox}{width=1\linewidth}
\begin{tabular}{c|c|llllll|lll}
\toprule
\multirow{2}{*}{Model} & \multicolumn{1}{c|}{Tasks} & \multicolumn{6}{c|}{Task1} & \multicolumn{3}{c}{Task2} \\
\cline{2-11}
 & \multicolumn{1}{c|}{Metrics} & Recall@20 & Recall@40 & NDCG@20 & NDCG@40 & Hit@20 & Hit@40 & PACA@10 & PACA@20 & PACA@40 \\
\hline
\rowcolor{skyblue}\multicolumn{11}{c}{\textit{Closed-source LLMs}}\\
\hline
\multirow{3}{*}{GPT-5} & w/o RAG & 0.061 &  0.090 &  0.263 &  0.212 &  0.115 &  0.176 &  0.053 &  0.069 & 0.081\\
 & RAG (R=5) & $0.065_{\textcolor{grey}{\uparrow 6.56\%}}$  & $0.093_{\textcolor{grey}{\uparrow 3.33\%}}$ & $0.306_{\textcolor{grey}{\uparrow 16.3\%}}$ & $0.231_{\textcolor{grey}{\uparrow 8.96\%}}$  & $0.124_{\textcolor{grey}{\uparrow 7.83\%}}$ & $0.181_{\textcolor{grey}{\uparrow 2.84\%}}$ & $0.086_{\textcolor{grey}{\uparrow 62.3\%}}$  & $0.109_{\textcolor{grey}{\uparrow 58.0\%}}$ & $0.131_{\textcolor{grey}{\uparrow 61.7\%}}$ \\
 & RAG (R=10) &  $0.071_{\textcolor{grey}{\uparrow 16.4\%}}$  & $\underline{0.099}_{\textcolor{grey}{\uparrow 10.0\%}}$ & $0.324_{\textcolor{grey}{\uparrow 23.2\%}}$ & $0.248_{\textcolor{grey}{\uparrow 17.0\%}}$  & $0.136_{\textcolor{grey}{\uparrow 18.3\%}}$ & $\underline{0.193}_{\textcolor{grey}{\uparrow 9.66\%}}$ & $0.098_{\textcolor{grey}{\uparrow 84.9\%}}$  & $0.121_{\textcolor{grey}{\uparrow 75.4\%}}$ & $0.143_{\textcolor{grey}{\uparrow 76.5\%}}$ \\
\hline
\multirow{3}{*}{GPT-o3} & w/o RAG & 0.029 & 0.034 & 0.166 & 0.113 & 0.059 & 0.068 & 0.030 & 0.036 & 0.041\\
 & RAG (R=5) &  $0.044_{\textcolor{grey}{\uparrow 51.7\%}}$ & $0.051_{\textcolor{grey}{\uparrow 50.0\%}}$  & $0.230_{\textcolor{grey}{\uparrow 38.6\%}}$ & $0.157_{\textcolor{grey}{\uparrow 38.9\%}}$   & $0.086_{\textcolor{grey}{\uparrow 45.8\%}}$ & $0.101_{\textcolor{grey}{\uparrow 48.5\%}}$ & $0.039_{\textcolor{grey}{\uparrow 30.0\%}}$  & $0.049_{\textcolor{grey}{\uparrow 36.1\%}}$ & $0.060_{\textcolor{grey}{\uparrow 46.3\%}}$ \\
 & RAG (R=10) &  $0.055_{\textcolor{grey}{\uparrow 89.7\%}}$  & $0.064_{\textcolor{grey}{\uparrow 88.2\%}}$ & $0.264_{\textcolor{grey}{\uparrow 59.0\%}}$ & $0.184_{\textcolor{grey}{\uparrow 62.8\%}}$  & $0.104_{\textcolor{grey}{\uparrow 76.3\%}}$ & $0.126_{\textcolor{grey}{\uparrow 85.3\%}}$ & $0.045_{\textcolor{grey}{\uparrow 50.0\%}}$  & $0.058_{\textcolor{grey}{\uparrow 61.1\%}}$ & $0.068_{\textcolor{grey}{\uparrow 65.9\%}}$ \\
\hline
\multirow{3}{*}{Claude-3.5} & w/o RAG & 0.028 & 0.034 & 0.140 & 0.098 & 0.055 & 0.067 & 0.024 & 0.029 & 0.034\\
 & RAG (R=5) & $0.044_{\textcolor{grey}{\uparrow 57.1\%}}$  & $0.050_{\textcolor{grey}{\uparrow 47.1\%}}$ & $0.230_{\textcolor{grey}{\uparrow 64.3\%}}$ & $0.157_{\textcolor{grey}{\uparrow 60.2\%}}$  & $0.084_{\textcolor{grey}{\uparrow 52.7\%}}$ & $0.098_{\textcolor{grey}{\uparrow 46.3\%}}$ & $0.046_{\textcolor{grey}{\uparrow 91.7\%}}$  & $0.056_{\textcolor{grey}{\uparrow 93.1\%}}$ & $0.064_{\textcolor{grey}{\uparrow 88.2\%}}$ \\
 & RAG (R=10) &  $0.054_{\textcolor{grey}{\uparrow 92.9\%}}$  & $0.062_{\textcolor{grey}{\uparrow 82.4\%}}$ & $0.264_{\textcolor{grey}{\uparrow 88.6\%}}$ & $0.180_{\textcolor{grey}{\uparrow 83.7\%}}$  & $0.103_{\textcolor{grey}{\uparrow 87.3\%}}$ & $0.121_{\textcolor{grey}{\uparrow 80.6\%}}$ & $0.057_{\textcolor{grey}{\uparrow 138\%}}$  & $0.069_{\textcolor{grey}{\uparrow 138\%}}$ & $0.078_{\textcolor{grey}{\uparrow 129\%}}$ \\
\hline
\multirow{3}{*}{Claude-3.7} & w/o RAG & 0.042 & 0.054 & 0.197 & 0.144 & 0.081 & 0.107 & 0.040 & 0.054 & 0.071 \\
 & RAG (R=5) &  $0.057_{\textcolor{grey}{\uparrow 35.7\%}}$  & $0.071_{\textcolor{grey}{\uparrow 31.5\%}}$ & $0.275_{\textcolor{grey}{\uparrow 39.6\%}}$ & $0.196_{\textcolor{grey}{\uparrow 36.1\%}}$  & $0.110_{\textcolor{grey}{\uparrow 35.8\%}}$ & $0.140_{\textcolor{grey}{\uparrow 30.8\%}}$ & $0.063_{\textcolor{grey}{\uparrow 57.5\%}}$  & $0.081_{\textcolor{grey}{\uparrow 50.0\%}}$ & $0.099_{\textcolor{grey}{\uparrow 39.4\%}}$ \\
 & RAG (R=10) &  $0.065_{\textcolor{grey}{\uparrow 54.8\%}}$  & $0.081_{\textcolor{grey}{\uparrow 50.0\%}}$ & $0.300_{\textcolor{grey}{\uparrow 52.3\%}}$ & $0.217_{\textcolor{grey}{\uparrow 50.7\%}}$  & $0.126_{\textcolor{grey}{\uparrow 55.6\%}}$ & $0.162_{\textcolor{grey}{\uparrow 51.4\%}}$ & $0.082_{\textcolor{grey}{\uparrow 105\%}}$  & $0.105_{\textcolor{grey}{\uparrow 94.4\%}}$ & $0.125_{\textcolor{grey}{\uparrow 76.1\%}}$ \\
\hline
\multirow{3}{*}{Claude-4-Sonnet} & w/o RAG & 0.038 & 0.051 & 0.187 & 0.139 & 0.076 & 0.103 & 0.046 & 0.064 & 0.080\\
 & RAG (R=5) & $0.054_{\textcolor{grey}{\uparrow 42.1\%}}$  & $0.068_{\textcolor{grey}{\uparrow 33.3\%}}$ & $0.268_{\textcolor{grey}{\uparrow 43.3\%}}$ & $0.193_{\textcolor{grey}{\uparrow 38.8\%}}$  & $0.100_{\textcolor{grey}{\uparrow 31.6\%}}$ & $0.136_{\textcolor{grey}{\uparrow 32.0\%}}$ & $0.076_{\textcolor{grey}{\uparrow 65.2\%}}$  & $0.097_{\textcolor{grey}{\uparrow 51.6\%}}$ & $0.115_{\textcolor{grey}{\uparrow 43.8\%}}$ \\
 & RAG (R=10) &  $0.064_{\textcolor{grey}{\uparrow 68.4\%}}$  & $0.080_{\textcolor{grey}{\uparrow 56.9\%}}$ & $0.296_{\textcolor{grey}{\uparrow 58.3\%}}$ & $0.213_{\textcolor{grey}{\uparrow 53.2\%}}$  & $0.122_{\textcolor{grey}{\uparrow 60.5\%}}$ & $0.157_{\textcolor{grey}{\uparrow 52.4\%}}$ & $0.080_{\textcolor{grey}{\uparrow 73.9\%}}$  & $0.101_{\textcolor{grey}{\uparrow 57.8\%}}$ & $0.122_{\textcolor{grey}{\uparrow 52.5\%}}$ \\
\hline
\multirow{3}{*}{Gemini-2.0-Flash} & w/o RAG & 0.023 & 0.030 & 0.115 & 0.084 & 0.045 & 0.060 & 0.028 & 0.039 & 0.051 \\
 & RAG (R=5) & $0.035_{\textcolor{grey}{\uparrow 52.2\%}}$  & $0.043_{\textcolor{grey}{\uparrow 43.3\%}}$ & $0.180_{\textcolor{grey}{\uparrow 56.5\%}}$ & $0.127_{\textcolor{grey}{\uparrow 51.2\%}}$  & $0.068_{\textcolor{grey}{\uparrow 51.1\%}}$ & $0.086_{\textcolor{grey}{\uparrow 43.3\%}}$ & $0.082_{\textcolor{grey}{\uparrow 193\%}}$  & $0.105_{\textcolor{grey}{\uparrow 169\%}}$ & $0.134_{\textcolor{grey}{\uparrow 163\%}}$ \\
 & RAG (R=10) &  $0.048_{\textcolor{grey}{\uparrow 109\%}}$  & $0.058_{\textcolor{grey}{\uparrow 93.3\%}}$ & $0.230_{\textcolor{grey}{\uparrow 100\%}}$ & $0.162_{\textcolor{grey}{\uparrow 92.9\%}}$  & $0.092_{\textcolor{grey}{\uparrow 104\%}}$ & $0.114_{\textcolor{grey}{\uparrow 90.0\%}}$ & $0.107_{\textcolor{grey}{\uparrow 282\%}}$  & $0.138_{\textcolor{grey}{\uparrow 254\%}}$ & $0.171_{\textcolor{grey}{\uparrow 235\%}}$ \\
\hline
\multirow{3}{*}{Gemini-2.5-Pro} & w/o RAG & 0.055 & 0.069 & 0.272 & 0.195 & 0.108 & 0.138 & 0.071 & 0.092 & 0.113\\
 & RAG (R=5) & $0.069_{\textcolor{grey}{\uparrow 25.5\%}}$  & $0.087_{\textcolor{grey}{\uparrow 26.1\%}}$ & $0.329_{\textcolor{grey}{\uparrow 21.0\%}}$ & $0.235_{\textcolor{grey}{\uparrow 20.5\%}}$  & $0.134_{\textcolor{grey}{\uparrow 24.1\%}}$ & $0.170_{\textcolor{grey}{\uparrow 23.2\%}}$ & $0.099_{\textcolor{grey}{\uparrow 39.4\%}}$  & $0.129_{\textcolor{grey}{\uparrow 40.2\%}}$ & $0.155_{\textcolor{grey}{\uparrow 37.2\%}}$ \\
 & RAG (R=10) &  $\underline{0.076}_{\textcolor{grey}{\uparrow 38.2\%}}$  & $0.097_{\textcolor{grey}{\uparrow 40.6\%}}$ & $\underline{0.346}_{\textcolor{grey}{\uparrow 27.2\%}}$ & $\underline{0.252}_{\textcolor{grey}{\uparrow 29.2\%}}$  & $\underline{0.144}_{\textcolor{grey}{\uparrow 33.3\%}}$ & $0.188_{\textcolor{grey}{\uparrow 36.2\%}}$ & 
 $\underline{0.127}_{\textcolor{grey}{\uparrow 78.9\%}}$  & $\underline{0.170}_{\textcolor{grey}{\uparrow 84.8\%}}$ & $\underline{0.201}_{\textcolor{grey}{\uparrow 77.9\%}}$ \\
\hline
\multirow{3}{*}{Grok-3} & w/o RAG & 0.016 & 0.017 & 0.090 & 0.060 & 0.031 & 0.034 & 0.030 & 0.043 & 0.054\\
 & RAG (R=5) & $0.024_{\textcolor{grey}{\uparrow 50.0\%}}$  & $0.025_{\textcolor{grey}{\uparrow 47.1\%}}$ & $0.134_{\textcolor{grey}{\uparrow 48.9\%}}$ & $0.087_{\textcolor{grey}{\uparrow 45.0\%}}$  & $0.047_{\textcolor{grey}{\uparrow 51.6\%}}$ & $0.049_{\textcolor{grey}{\uparrow 44.1\%}}$ & $0.082_{\textcolor{grey}{\uparrow 173\%}}$  & $0.092_{\textcolor{grey}{\uparrow 114\%}}$ & $0.098_{\textcolor{grey}{\uparrow 81.5\%}}$ \\
 & RAG (R=10) &  $0.040_{\textcolor{grey}{\uparrow 150\%}}$  & $0.042_{\textcolor{grey}{\uparrow 147\%}}$ & $0.192_{\textcolor{grey}{\uparrow 113\%}}$ & $0.122_{\textcolor{grey}{\uparrow 103\%}}$  & $0.077_{\textcolor{grey}{\uparrow 148\%}}$ & $0.079_{\textcolor{grey}{\uparrow 132\%}}$ & $0.101_{\textcolor{grey}{\uparrow 237\%}}$  & $0.122_{\textcolor{grey}{\uparrow 184\%}}$ & $0.133_{\textcolor{grey}{\uparrow 146\%}}$ \\
\hline
\rowcolor{skyblue}\multicolumn{11}{c}{\textit{Open-source LLMs}}\\
\hline
\multirow{3}{*}{Kimi-K2} & w/o RAG & 0.036 & 0.046 & 0.177 & 0.127 & 0.071 & 0.091 & 0.048 & 0.066 & 0.080 \\
 & RAG (R=5) & $0.052_{\textcolor{grey}{\uparrow 44.4\%}}$  & $0.064_{\textcolor{grey}{\uparrow 39.1\%}}$ & $0.245_{\textcolor{grey}{\uparrow 38.4\%}}$ & $0.175_{\textcolor{grey}{\uparrow 37.8\%}}$  & $0.100_{\textcolor{grey}{\uparrow 40.8\%}}$ & $0.126_{\textcolor{grey}{\uparrow 38.5\%}}$ & $0.085_{\textcolor{grey}{\uparrow 77.1\%}}$  & $0.104_{\textcolor{grey}{\uparrow 57.6\%}}$ & $0.120_{\textcolor{grey}{\uparrow 50.0\%}}$ \\
 & RAG (R=10) &  $0.061_{\textcolor{grey}{\uparrow 69.4\%}}$  & $0.075_{\textcolor{grey}{\uparrow 63.0\%}}$ & $0.273_{\textcolor{grey}{\uparrow 54.2\%}}$ & $0.195_{\textcolor{grey}{\uparrow 53.5\%}}$  & $0.117_{\textcolor{grey}{\uparrow 64.8\%}}$ & $0.146_{\textcolor{grey}{\uparrow 60.4\%}}$ & $0.098_{\textcolor{grey}{\uparrow 104\%}}$  & $0.123_{\textcolor{grey}{\uparrow 86.4\%}}$ & $0.143_{\textcolor{grey}{\uparrow 78.7\%}}$ \\
\hline
\multirow{3}{*}{Deepseek-v3} & w/o RAG & 0.028 & 0.033 & 0.143 & 0.099 & 0.055 & 0.067 & 0.048 & 0.063 & 0.073 \\
 & RAG (R=5) & $0.050_{\textcolor{grey}{\uparrow 78.6\%}}$  & $0.058_{\textcolor{grey}{\uparrow 75.8\%}}$ & $0.248_{\textcolor{grey}{\uparrow 73.4\%}}$ & $0.171_{\textcolor{grey}{\uparrow 72.7\%}}$  & $0.096_{\textcolor{grey}{\uparrow 74.5\%}}$ & $0.114_{\textcolor{grey}{\uparrow 70.1\%}}$ & $0.103_{\textcolor{grey}{\uparrow 115\%}}$  & $0.131_{\textcolor{grey}{\uparrow 108\%}}$ & $0.153_{\textcolor{grey}{\uparrow 110\%}}$ \\
 & RAG (R=10) &  $0.060_{\textcolor{grey}{\uparrow 114\%}}$  & $0.070_{\textcolor{grey}{\uparrow 112\%}}$ & $0.278_{\textcolor{grey}{\uparrow 94.4\%}}$ & $0.193_{\textcolor{grey}{\uparrow 94.9\%}}$  & $0.111_{\textcolor{grey}{\uparrow 102\%}}$ & $0.137_{\textcolor{grey}{\uparrow 104\%}}$ & $0.108_{\textcolor{grey}{\uparrow 125\%}}$  & $0.135_{\textcolor{grey}{\uparrow 114\%}}$ & $0.158_{\textcolor{grey}{\uparrow 116\%}}$ \\
\hline
\multirow{3}{*}{Qwen3-Coder-480B} & w/o RAG & 0.019 & 0.023 & 0.100 & 0.071 & 0.037 & 0.047 & 0.042 & 0.053 & 0.065 \\
 & RAG (R=5) & $0.037_{\textcolor{grey}{\uparrow 94.7\%}}$  & $0.042_{\textcolor{grey}{\uparrow 82.6\%}}$ & $0.177_{\textcolor{grey}{\uparrow 77.0\%}}$ & $0.121_{\textcolor{grey}{\uparrow 70.4\%}}$  & $0.072_{\textcolor{grey}{\uparrow 94.6\%}}$ & $0.083_{\textcolor{grey}{\uparrow 76.6\%}}$ & $0.101_{\textcolor{grey}{\uparrow 140\%}}$  & $0.121_{\textcolor{grey}{\uparrow 128\%}}$ & $0.138_{\textcolor{grey}{\uparrow 112\%}}$ \\
 & RAG (R=10) &  $0.049_{\textcolor{grey}{\uparrow 158\%}}$  & $0.054_{\textcolor{grey}{\uparrow 135\%}}$ & $0.216_{\textcolor{grey}{\uparrow 116\%}}$ & $0.147_{\textcolor{grey}{\uparrow 107\%}}$  & $0.095_{\textcolor{grey}{\uparrow 157\%}}$ & $0.106_{\textcolor{grey}{\uparrow 126\%}}$ & $0.113_{\textcolor{grey}{\uparrow 169\%}}$  & $0.140_{\textcolor{grey}{\uparrow 164\%}}$ & $0.161_{\textcolor{grey}{\uparrow 148\%}}$ \\
\hline
\multirow{3}{*}{GPT-oss-120B} & w/o RAG & 0.016 & 0.018 & 0.095 & 0.063 & 0.033 & 0.037 & 0.024 & 0.028 & 0.031 \\
 & RAG (R=5) & $0.027_{\textcolor{grey}{\uparrow 68.8\%}}$  & $0.031_{\textcolor{grey}{\uparrow 72.2\%}}$ & $0.155_{\textcolor{grey}{\uparrow 63.2\%}}$ & $0.103_{\textcolor{grey}{\uparrow 63.5\%}}$  & $0.054_{\textcolor{grey}{\uparrow 63.6\%}}$ & $0.060_{\textcolor{grey}{\uparrow 62.2\%}}$ & $0.069_{\textcolor{grey}{\uparrow 188\%}}$  & $0.082_{\textcolor{grey}{\uparrow 193\%}}$ & $0.090_{\textcolor{grey}{\uparrow 190\%}}$ \\
 & RAG (R=10) &  $0.043_{\textcolor{grey}{\uparrow 169\%}}$  & $0.046_{\textcolor{grey}{\uparrow 156\%}}$ & $0.219_{\textcolor{grey}{\uparrow 131\%}}$ & $0.144_{\textcolor{grey}{\uparrow 129\%}}$  & $0.083_{\textcolor{grey}{\uparrow 152\%}}$ & $0.090_{\textcolor{grey}{\uparrow 143\%}}$ & $0.087_{\textcolor{grey}{\uparrow 263\%}}$  & $0.108_{\textcolor{grey}{\uparrow 286\%}}$ & $0.122_{\textcolor{grey}{\uparrow 294\%}}$ \\
\hline
\multirow{6}{*}{CitationGenerator-4B} & w/o All & 0.006 & 0.007 & 0.029 & 0.021 & 0.013 & 0.017 & 0.009 & 0.010 & 0.012 \\
 & w/o SFT (R=5)  & $0.011_{\textcolor{grey}{\uparrow 83.3\%}}$  & $0.012_{\textcolor{grey}{\uparrow 71.4\%}}$ & $0.056_{\textcolor{grey}{\uparrow 93.1\%}}$ & $0.034_{\textcolor{grey}{\uparrow 61.9\%}}$  & $0.024_{\textcolor{grey}{\uparrow 84.6\%}}$ & $0.027_{\textcolor{grey}{\uparrow 58.8\%}}$ & $0.051_{\textcolor{grey}{\uparrow 467\%}}$  & $0.064_{\textcolor{grey}{\uparrow 540\%}}$ & $0.074_{\textcolor{grey}{\uparrow 517\%}}$ \\
 & w/o SFT (R=10)  & $0.021_{\textcolor{grey}{\uparrow 250\%}}$  & $0.023_{\textcolor{grey}{\uparrow 229\%}}$ & $0.122_{\textcolor{grey}{\uparrow 321\%}}$ & $0.078_{\textcolor{grey}{\uparrow 271\%}}$  & $0.041_{\textcolor{grey}{\uparrow 215\%}}$ & $0.045_{\textcolor{grey}{\uparrow 165\%}}$ & $0.063_{\textcolor{grey}{\uparrow 600\%}}$  & $0.078_{\textcolor{grey}{\uparrow 680\%}}$ & $0.092_{\textcolor{grey}{\uparrow 667\%}}$ \\
 & w/o RAG & 0.012 & 0.019 & 0.055 & 0.047 & 0.022 & 0.039 & 0.017 & 0.029 & 0.042 \\
 & Full (R=5)  & $0.023_{\textcolor{grey}{\uparrow 91.7\%}}$  & $0.034_{\textcolor{grey}{\uparrow 78.9\%}}$ & $0.107_{\textcolor{grey}{\uparrow 94.5\%}}$ & $0.084_{\textcolor{grey}{\uparrow 78.7\%}}$  & $0.042_{\textcolor{grey}{\uparrow 90.9\%}}$ & $0.065_{\textcolor{grey}{\uparrow 66.7\%}}$ & $0.087_{\textcolor{grey}{\uparrow 412\%}}$  & $0.096_{\textcolor{grey}{\uparrow 231\%}}$ & $0.101_{\textcolor{grey}{\uparrow 140\%}}$ \\
 & Full (R=10)  & $0.027_{\textcolor{grey}{\uparrow 125\%}}$  & $0.039_{\textcolor{grey}{\uparrow 105\%}}$ & $0.129_{\textcolor{grey}{\uparrow 135\%}}$ & $0.098_{\textcolor{grey}{\uparrow 109\%}}$  & $0.053_{\textcolor{grey}{\uparrow 141\%}}$ & $0.074_{\textcolor{grey}{\uparrow 89.7\%}}$ & $0.119_{\textcolor{grey}{\uparrow 600\%}}$  & $0.148_{\textcolor{grey}{\uparrow 410\%}}$ & $0.164_{\textcolor{grey}{\uparrow 290\%}}$ \\
 \hline
\multirow{6}{*}{CitationGenerator-30B} & w/o All &  0.010 & 0.011 & 0.057 & 0.038 & 0.020 & 0.023 & 0.014 & 0.016 & 0.017 \\
 & w/o SFT (R=5) & $0.021_{\textcolor{grey}{\uparrow 110\%}}$  & $0.024_{\textcolor{grey}{\uparrow 118\%}}$ & $0.117_{\textcolor{grey}{\uparrow 105\%}}$ & $0.076_{\textcolor{grey}{\uparrow 100\%}}$  & $0.039_{\textcolor{grey}{\uparrow 95.0\%}}$ & $0.042_{\textcolor{grey}{\uparrow 82.6\%}}$ & $0.076_{\textcolor{grey}{\uparrow 443\%}}$  & $0.084_{\textcolor{grey}{\uparrow 425\%}}$ & $0.089_{\textcolor{grey}{\uparrow 424\%}}$ \\
 & w/o SFT (R=10) & $0.031_{\textcolor{grey}{\uparrow 210\%}}$  & $0.035_{\textcolor{grey}{\uparrow 218\%}}$ & $0.153_{\textcolor{grey}{\uparrow 168\%}}$ & $0.100_{\textcolor{grey}{\uparrow 163\%}}$  & $0.059_{\textcolor{grey}{\uparrow 195\%}}$ & $0.063_{\textcolor{grey}{\uparrow 174\%}}$ & $0.096_{\textcolor{grey}{\uparrow 586\%}}$  & $0.114_{\textcolor{grey}{\uparrow 613\%}}$ & $0.124_{\textcolor{grey}{\uparrow 629\%}}$ \\
 & w/o RAG &  0.043 & 0.069 & 0.225 & 0.198 & 0.113 & 0.130 & 0.149 & 0.165 & 0.191 \\
 & Full (R=5) & $0.072_{\textcolor{grey}{\uparrow 67.4\%}}$  & $0.113_{\textcolor{grey}{\uparrow 63.8\%}}$ & $0.338_{\textcolor{grey}{\uparrow 50.2\%}}$ & $0.273_{\textcolor{grey}{\uparrow 37.9\%}}$  & $0.183_{\textcolor{grey}{\uparrow 61.9\%}}$ & $0.204_{\textcolor{grey}{\uparrow 56.9\%}}$ & $0.268_{\textcolor{grey}{\uparrow 79.9\%}}$  & $0.291_{\textcolor{grey}{\uparrow 76.4\%}}$ & $0.318_{\textcolor{grey}{\uparrow 66.5\%}}$ \\
 & Full (R=10) & $\mathbf{0.076}^*_{\textcolor{grey}{\uparrow 76.7\%}}$  & $\mathbf{0.122}^*_{\textcolor{grey}{\uparrow 76.8\%}}$ & $\mathbf{0.367}^*_{\textcolor{grey}{\uparrow 63.1\%}}$ & $\mathbf{0.295}^*_{\textcolor{grey}{\uparrow 49.0\%}}$  & $\mathbf{0.193}^*_{\textcolor{grey}{\uparrow 70.8\%}}$ & $\mathbf{0.219}^*_{\textcolor{grey}{\uparrow 68.5\%}}$ & $\mathbf{0.273}^*_{\textcolor{grey}{\uparrow 83.2\%}}$  & $\mathbf{0.303}^*_{\textcolor{grey}{\uparrow 83.6\%}}$ & $\mathbf{0.342}^*_{\textcolor{grey}{\uparrow 79.1\%}}$ \\

\bottomrule
\end{tabular}
\end{adjustbox}
\vspace{-1mm}
\label{tab:overall}
\end{table*}

\subsection{Baselines}

\subsubsection{Closed-source LLMs}

We evaluate leading commercial language models from major AI companies, including OpenAI (GPT-o3 and GPT-5), Anthropic (Claude-3.5, Claude-3.7, and Claude Sonnet 4), Google (Gemini-2.5-Pro and Gemini-2.0-Flash), and xAI (Grok-3). 
\subsubsection{Open-source LLMs}

The open-source evaluation encompasses prominent publicly available models, including Moonshot AI (Kimi-K2), DeepSeek (DeepSeek-V3), OpenAI (GPT-oss-120B), and Qwen (Qwen3-Coder-480B). 
Additionally, we employ the Qwen3 series as our primary experimental models CitationGenerator to demonstrate specialized model adaptation for academic citation prediction tasks. 

\subsection{Experimental Settings}

We conduct the following fourfold experiments: (1) \textbf{Few-shot In-context Learning:} Leverages large language models' contextual understanding without parameter updates by constructing prompts with representative training examples, task-specific instructions, and JSON output formats, emphasizing holistic paper comprehension for List-Specific Citation Prediction and contextual analysis around reference placeholders for Position-Specific Citation Prediction, while incorporating reasoning to prevent hallucination and ensure grounded predictions.  
(2) \textbf{Retrieval Augmented Generation:} Integrates external knowledge retrieval with generation in a two-stage pipeline, querying a hierarchical corpus using paper titles and abstracts to return top-R relevant papers (R=5 or R=10) based on semantic similarity, which are then incorporated into generation prompts alongside target paper content to ground predictions in actual corpus content.  
(3) \textbf{Supervised Fine-tuning:} Adapts pre-trained language models for citation prediction through parameter optimization on task-specific data, using question-answer formats with paper metadata as input and citation predictions in JSON schemas as targets, with quality-filtered and consistently formatted training data to ensure stable convergence.  
(4) \textbf{Supervised Fine-tuning with Retrieval Augmented Generation:} Combines parametric adaptation with external knowledge access by first fine-tuning on citation-specific tasks and then integrating retrieved papers during training and inference, enabling models to leverage both adapted parametric knowledge and external corpus content.

\subsection{Implementation Details}
Our retriever module was fine-tuned from the Qwen3-Embedding-8B~\cite{qwen3-emb} model using the Swift framework. The training was configured with a learning rate of 5e-5, a batch size of 16, and a maximum sequence length of 4,608. For the generator, we fine-tuned the Qwen3-4B~\cite{qwen3} and Qwen3-Coder-30B~\cite{qwen25coder} models utilizing the Megatron-LM~\cite{megatron} framework. These models were trained with a learning rate of 7e-6, a batch size of 64, and a maximum sequence length of 131,072.

For inference, all open-source models were served using vLLM, while closed-source models were queried via their official APIs. To ensure deterministic and focused outputs, we set the generation temperature to 0.1 and the presence penalty to 1.0 for all model requests, wherever supported by the respective API\footnote{GPT-o3 does not support custom temperature or max\_tokens settings. Claude models do not support presence penalty configuration.}. All training and deployment were conducted on a cluster of 32 NVIDIA A100 GPUs.

\section{Experimental Results}
\label{sec:results}
\subsection{Results on Task 1 and Task 2}

\subsubsection{Overall Ratings}

Table~\ref{tab:overall} presents comprehensive performance comparisons across both citation prediction tasks. RAG consistently improves performance across all models, with retrieval depth significantly impacting prediction quality. Increasing retrieval from R=5 to R=10 yields substantial improvements, with models like Grok-3 and Claude-3.5 showing over 80\% relative improvement in Task 1 Recall@20. Task 2 exhibits even more pronounced benefits, with Gemini-2.0-Flash achieving over 200\% improvement in PACA metrics with RAG integration. Supervised fine-tuning demonstrates remarkable effectiveness, as CitationGenerator-30B without RAG achieves competitive performance with leading closed-source models that use retrieval, particularly excelling in Task 2 with PACA@20 of 0.165. The combination of fine-tuning and RAG produces synergistic effects, with CitationGenerator-30B (R=10) achieving the highest performance: Recall@20 of 0.076, NDCG@20 of 0.367 for Task 1, and PACA@20 of 0.303 for Task 2. Baseline capabilities vary considerably, with closed-source models demonstrating stronger zero-shot performance, though this gap narrows substantially with domain-specific fine-tuning.

\subsubsection{Ablation Study}

Table~\ref{tab:overall} demonstrates that supervised fine-tuning and retrieval augmentation contribute through distinct yet complementary mechanisms in our CitationGenerator models. Fine-tuning provides the primary performance gain, with w/o RAG configurations substantially outperforming w/o SFT variants across both model sizes, demonstrating that domain adaptation is essential for capturing academic citation patterns. Retrieval augmentation further enhances fine-tuned models, with Full configurations achieving significant improvements over w/o RAG baselines. Model capacity plays a critical role, as CitationGenerator-4B shows limited effectiveness across all configurations while CitationGenerator-30B achieves state-of-the-art performance, highlighting the importance of sufficient parameter scale for complex citation prediction tasks.

\subsection{Retriever Comparison}

\begin{table}[!t]
\caption{Results on retriever comparison of different models.}
\centering
\begin{adjustbox}{width=1\linewidth}
\begin{tabular}{lcccc}
\toprule
\textbf{Model} & \textbf{Recall@20} & \textbf{Recall@50} & \textbf{MRR@20} & \textbf{MRR@50} \\
\midrule
TF-IDF & 0.0212 & 0.0332 & 0.1109 & 0.1145 \\
BM25 & 0.0283 & 0.0426 & 0.1402 & 0.1435 \\
BGE-M3 & 0.0326 & 0.0481 & 0.1611 & 0.1643 \\
All-MPNet-Base-V2 & 0.0366 & 0.0566 & 0.1670 & 0.1703 \\
Multilingual-E5-Large & 0.0332 & 0.0496 & 0.1634 & 0.1667 \\
Qwen3-Embedding-8B & 0.0522 & 0.0775 & 0.2163 & 0.2186 \\
CitationRetriever-8B & \textbf{0.0970} & \textbf{0.1386} & \textbf{0.3219} & \textbf{0.3232} \\
\bottomrule
\end{tabular}
\end{adjustbox}
\label{tab:retriever}
\end{table}

\subsubsection{Retriever Performance Comparison}

Table~\ref{tab:retriever} demonstrates substantial performance advantages of neural embedding models over traditional sparse retrieval methods for academic citation prediction. Traditional approaches including TF-IDF and BM25 achieve limited effectiveness with MRR@50 of 0.1145 and 0.1435. 
General-purpose neural models including BGE-M3 and All-MPNet-Base-V2 show moderate improvements with MRR@50 around 0.16-0.17, yet remain insufficient for citation-specific patterns. Qwen3-Embedding-8B exhibits stronger performance with MRR@50 of 0.2186 due to large-scale pre-training, though it lacks specialized adaptation for citation relationships. Our CitationRetriever-8B achieves MRR@50 of 0.3232, representing 47.9\% improvement over Qwen3-Embedding-8B and nearly double the performance of general-purpose models, demonstrating that contrastive learning on citation data enables capturing complex relationships including shared methodologies and field-specific conventions.

\subsubsection{Ablation on Multi-level Corpus}

\begin{figure}[!t]
    \centering
    \includegraphics[width=1.0\linewidth]{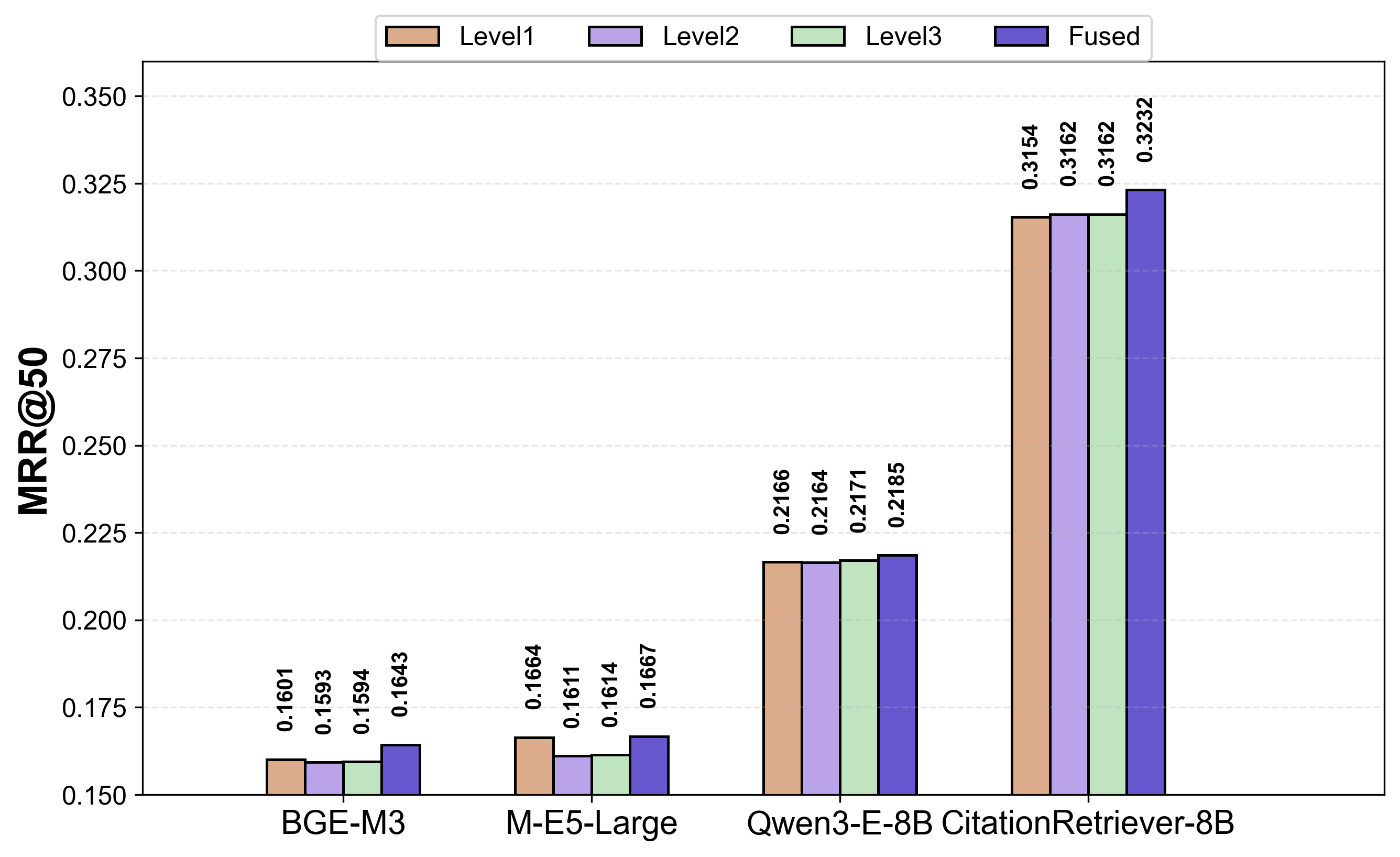}
        \vspace{-3mm}
    \caption{Performance Comparison of Single-Level vs. Multi-Level Fused Retrieval Strategies}
        \vspace{-3mm}
    \label{fig:placeholder}
\end{figure}

Figure~\ref{fig:placeholder} illustrates that multi-level fusion consistently outperforms single-level retrieval across all embedding models.
Traditional models show modest improvements of 0.2-2.7\% in MRR@50, while Qwen3-Embedding-8B demonstrates 1.1\% improvement. CitationRetriever-8B exhibits the most substantial gains with 3.7\% improvement over its best single-level configuration, indicating that contrastive fine-tuning enables learning complementary relevance signals across granularities where Level 1 captures topical alignment, Level 2 incorporates methodological context, and Level 3 provides comprehensive content coverage. These results validate that academic papers require multi-level representation for effective citation prediction.

\section{Related Work}

\subsection{Retrieval-Augmented Generation}

Traditional retrieval methods~\cite{zheng2025balancing,zheng-etal-2025-lagcl4rec,2025zhengnegative} are no longer sufficient to meet the demands of the era of large language models~\cite{li2025capgeo,feng2025seeing,zhang2025plotcraft,team2025qwen3,qwen25coder,megaflow,hangzhang25,kang2026quanteval,he2026order,zhang2026completion,wang2025audio}, retrieval-augmented generation (RAG) has become a key paradigm for mitigating large language models’ limitations in domain-specific or time-sensitive tasks, where models often hallucinate or generate outdated responses \cite{tonmoy2024comprehensive}. RAG addresses these limitations via a three-stage pipeline: retrieval from external knowledge bases, query–context integration, and grounded response generation \cite{asai2024self}.
Subsequent work improves RAG through enhanced retrievers \cite{lewis2020retrieval}, adaptive retrieval strategies \cite{jeong2024adaptive}, reasoning-based prompting such as IR-CoT \cite{trivedi2022interleaving}, and graph-structured retrieval like GraphRAG \cite{edge2024local}. However, these general frameworks lack the hierarchical corpus organization and multi-granularity retrieval needed for academic literature, where citation reasoning spans titles, abstracts, and full texts.

\subsection{Citation Generation and Evaluation}

Recent RAG systems have enabled citation-aware text generation for verifiability. Early models such as WebGPT \cite{nakano2021webgpt} and RL-based methods \cite{menick2022teaching} laid the groundwork, while later systems including ReGen \cite{qian2023webbrain} and WebCPM \cite{qin2023webcpm} refined citation integration through improved training. Recent advances in few-shot learning, self-reflection mechanisms, and fine-grained reward modeling have further improved citation quality. Evaluation has evolved from manual annotation (AIS \cite{rashkin2023measuring}) to automated and benchmark-based approaches such as ALCE \cite{gao2023enabling}. However, existing work mainly targets Web-based question answering and lacks support for academic citation prediction, which requires both document-level reference modeling and position-specific citation reasoning. 

\section{Conclusion}
In this paper, we propose \textbf{CiteRAG}, the first comprehensive RAG-integrated benchmark for evaluating large language models on academic citation prediction. The benchmark features dual-granularity tasks, a hierarchical corpus of 554k papers, and standardized evaluation framework. Extensive experiments across state-of-the-art language models demonstrate that RAG integration consistently enhances prediction accuracy and reduces hallucination rates. Our results reveal that specialized domain adaptation through supervised fine-tuning provides substantial performance gains, while retrieval augmentation offers complementary benefits through external knowledge grounding. Our open-source toolkit provides the research community with reproducible evaluation protocols and comprehensive baselines, establishing a methodological foundation for advancing citation prediction systems.

\section*{Acknowledgments}

This work is supported in part by the National Natural Science Foundation of China (No. 62372264 and No. 92467203) and Sina Weibo Corp. Chaokun Wang is the corresponding author.

\clearpage
\newpage
\balance
\bibliographystyle{ACM-Reference-Format}
\bibliography{sample-base}

\appendix
\section{QA Format}
\label{appendix:QA}
\begin{figure*}[!t]
    \centering
    \vspace{-1mm}
    \includegraphics[width=1.0\linewidth]{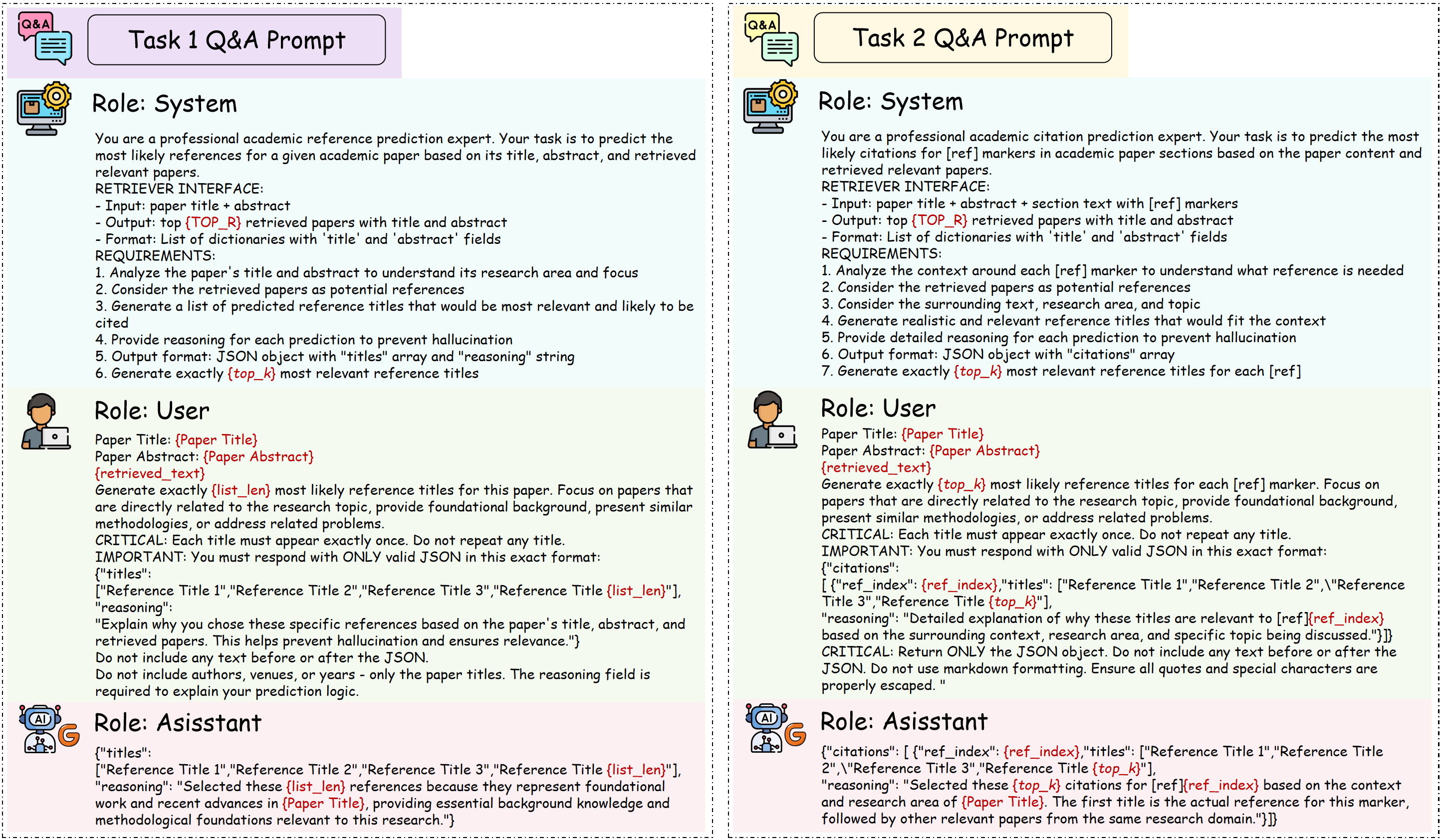}
    \vspace{-6mm}
\caption{Question-answer format specifications for Task 1 and Task 2.}
\vspace{-3mm}
\label{fig:QA}
\end{figure*}

Both citation prediction tasks are formatted as question-answer pairs following the ChatML format. The System role defines task requirements, specifying input format (paper metadata and retrieved context) and output format (JSON object with predicted titles and reasoning). The User role provides the query paper's title, abstract, and retrieved papers from the corpus, with Task 2 additionally including section text containing reference placeholders. The Assistant role generates predictions in strict JSON format, producing a ranked list of reference titles for Task 1 or position-specific citations for Task 2, accompanied by detailed reasoning to ground predictions in the provided context and prevent hallucination. Figure~\ref{fig:QA} illustrates the complete ChatML prompt structure and expected response format for both tasks.

\section{Detailed Metric Definitions}
\label{appendix:standard_metrics}
For the retriever evaluation, \textbf{Recall@k} measures the proportion of relevant papers successfully retrieved within the top-$k$ positions:
\begin{equation}
\text{Recall@k} = \frac{|\text{Retrieved@k} \cap \text{Relevant}|}{|\text{Relevant}|}
\end{equation}

For the retriever evaluation, \textbf{Mean Reciprocal Rank@k (MRR@k)} evaluates the ranking quality by considering the position of the first relevant item:
\begin{equation}
\text{MRR@k} = \frac{1}{|Q|} \sum_{i=1}^{|Q|} \frac{1}{\text{rank}_i}
\end{equation}
where ranks beyond k contribute zero to the average.

For the task 1 evaluation, \textbf{Recall@k} measures the proportion of ground truth references successfully predicted:
\begin{equation}
\text{Recall@k} = \frac{|\text{Predicted@k} \cap \text{GroundTruth}|}{|\text{GroundTruth}|}
\end{equation}

For the task 1 evaluation, \textbf{Normalized Discounted Cumulative Gain@k (NDCG@k)} evaluates both relevance and ranking quality:
\begin{equation}
\text{NDCG@k} = \frac{\text{DCG@k}}{\text{IDCG@k}}
\end{equation}
where
\begin{equation}
\text{DCG@k} = \sum_{i=1}^{k} \frac{2^{\text{rel}_i} - 1}{\log_2(i + 1)}
\end{equation}

For the task 1 evaluation, \textbf{Hit@k} quantifies the absolute number of correct predictions within the top-$k$ results.

\section{External Experimental Results}

\subsection{Citation Diversity and Hallucination Analysis}

\begin{figure}[!t]
    \centering
    \includegraphics[width=1.0\linewidth]{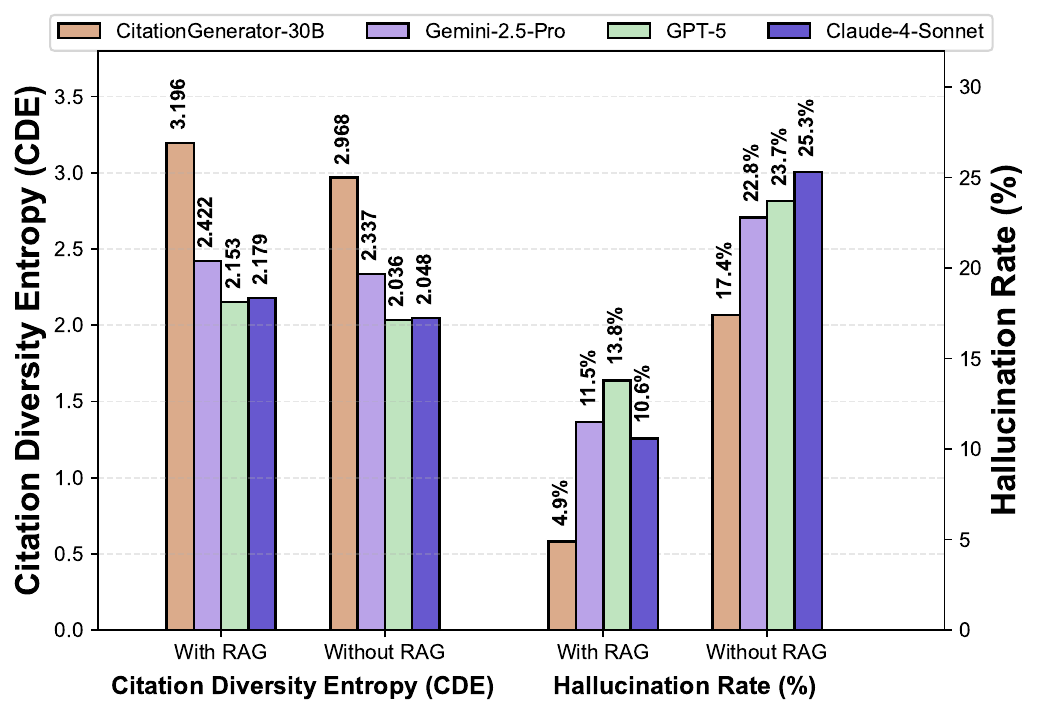}
    \vspace{-6mm}
    \caption{Citation Quality Assessment: Diversity Entropy and Hallucination Rate Across Models.}
    \vspace{-3mm}
    \label{fig:HD}
\end{figure}

Figure~\ref{fig:HD} reveals the complementary effects of RAG on citation diversity and hallucination reduction. RAG integration consistently increases citation diversity entropy, with CitationGenerator-30B achieving 3.196 with RAG compared to 2.968 without, indicating broader subcategory exploration. Closed-source models exhibit moderate diversity improvements of 0.3-0.5 entropy units while maintaining balanced distributions. More critically, RAG dramatically reduces hallucination rates, with CitationGenerator-30B decreasing from 17.4\% to 4.9\%, and Gemini-2.5-Pro, GPT-5, and Claude-4-Sonnet reducing from 22.8\%, 23.7\%, and 25.9\% to 11.5\%, 13.8\%, and 10.6\% respectively. 

\subsection{Retrieval Depth vs. Quality Trade-off Analysis}

\begin{figure}[!t]
\centering
\vspace{-1mm}
\begin{adjustbox}{width=1\linewidth}
\begin{tabular}{cc}
\includegraphics[width=0.49\textwidth]{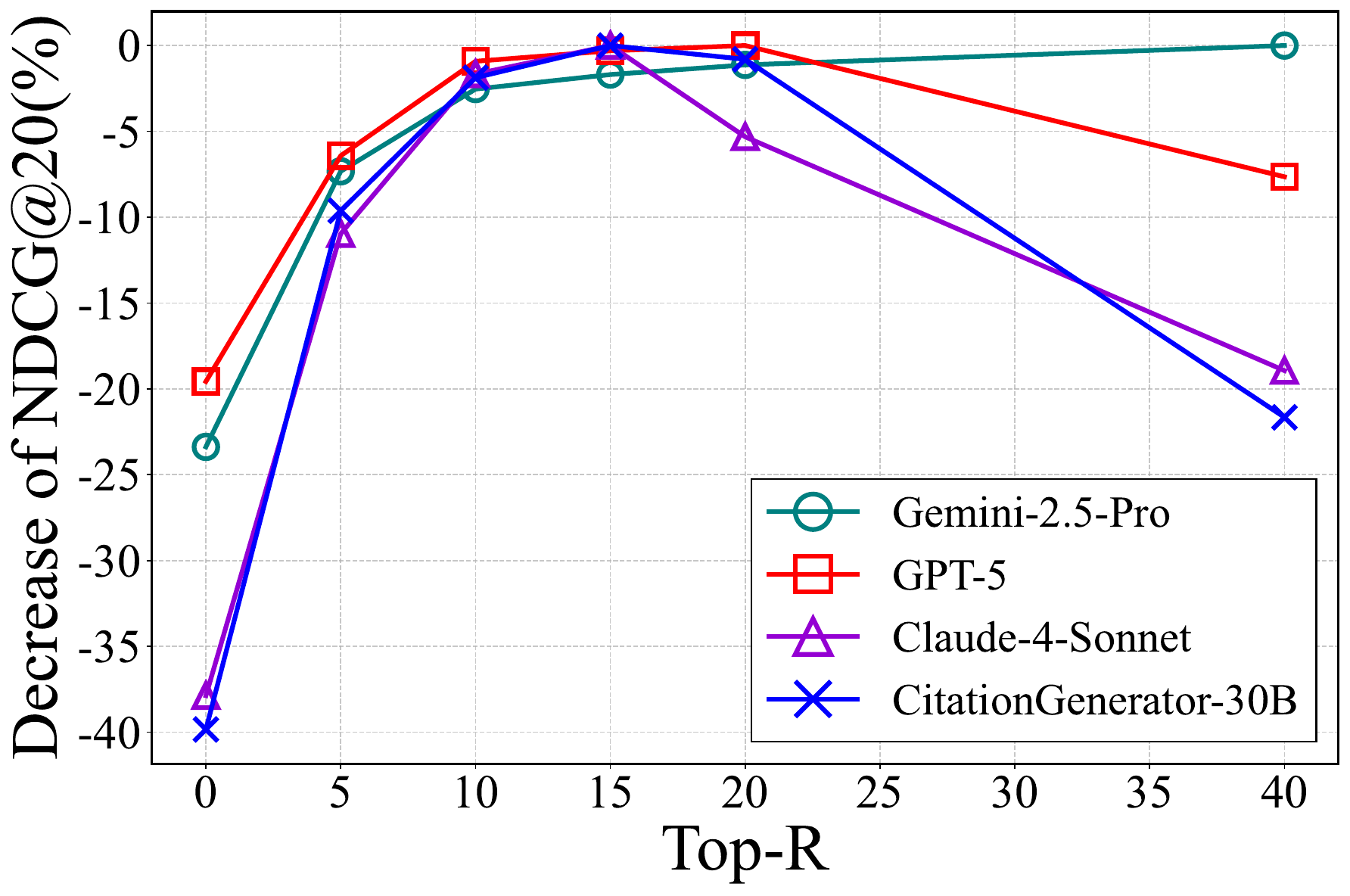} &
\includegraphics[width=0.49\textwidth]{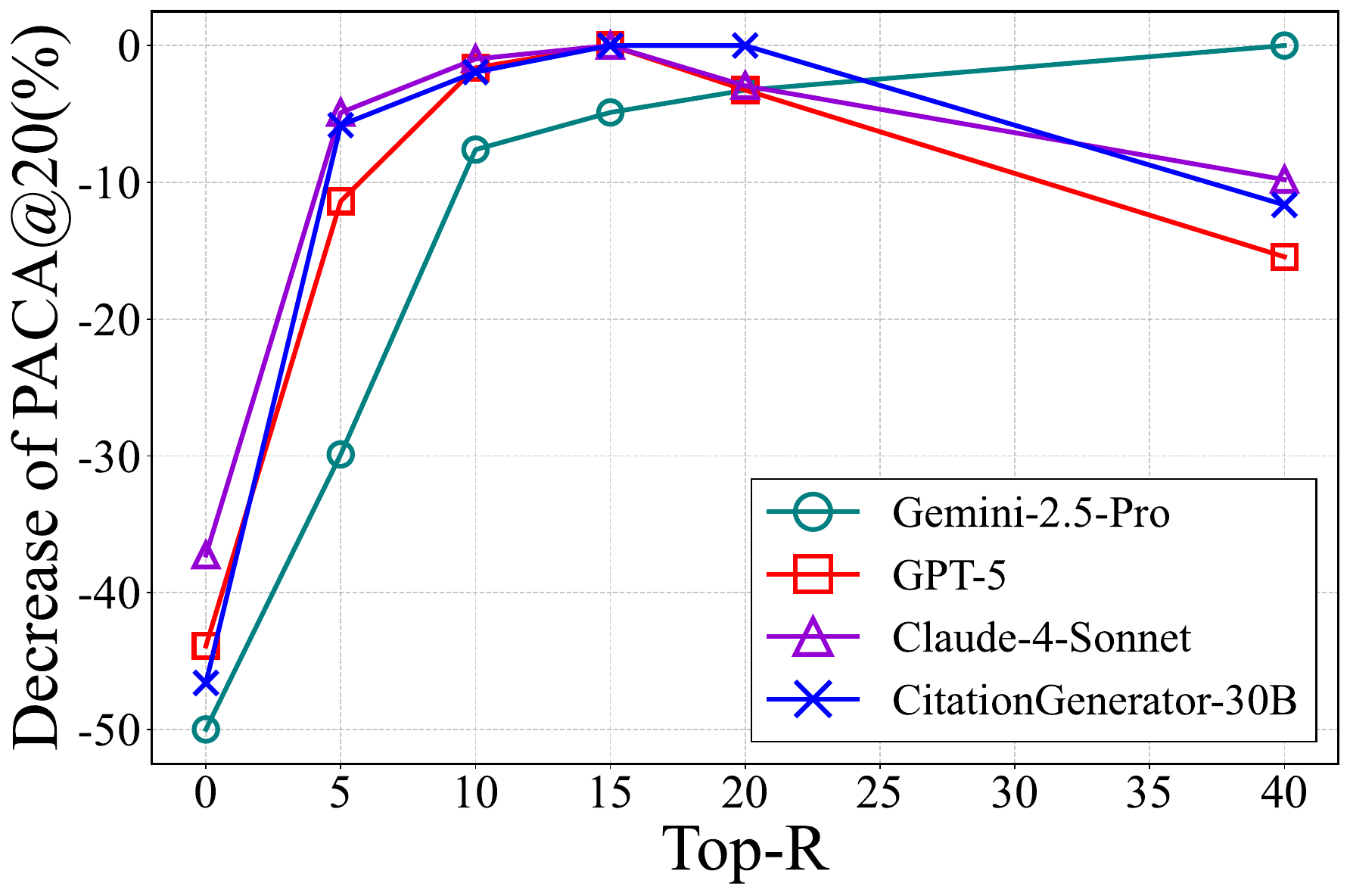} \\
\end{tabular}
\end{adjustbox}
\vspace{-3mm}
\caption{Retrieval Depth vs. Prediction Quality Trade-off Analysis.}
\vspace{-3mm}
\label{fig:efficiency}
\end{figure}

Figure~\ref{fig:efficiency} demonstrates complex relationships between retrieval depth and prediction quality influenced by context processing capabilities. Most models exhibit initial improvements from R=5-15, followed by performance degradation beyond optimal depth as excessive context introduces noise and exceeds attention span limits. Notably, Gemini-2.5-Pro maintains consistent performance improvement across all retrieval depths, attributed to its extended context window capability that effectively processes longer sequences without quality degradation. CitationGenerator-30B demonstrates competitive robustness within moderate ranges but shows limitations with extensive context. Considering the trade-offs between inference latency, token consumption costs, and prediction accuracy, R=10 emerges as the optimal configuration, providing substantial quality improvements over shallow retrieval while maintaining computational efficiency and avoiding the noise interference observed at deeper retrieval levels across most model architectures.

\subsection{Noise Robustness Analysis}
\label{app:noise}

\begin{table}[htbp]
\caption{Results on noise robustness comparison of different models.}
\centering
\begin{adjustbox}{width=1\linewidth}
\begin{tabular}{c|c|cc|c}
\toprule
\multirow{2}{*}{Model} & {Tasks} & \multicolumn{2}{c|}{Task1} & \multicolumn{1}{c}{Task2} \\
\cline{2-5}
 & {Metrics} & Recall@20 & NDCG@20 & PACA@20  \\
\hline
\multirow{6}{*}{GPT-5}
& w/o RAG & 0.061 & 0.263 & 0.069 \\
& Noise 0\% & 0.071 & 0.324 & 0.121 \\
 & Noise 20\% & 0.066 & 0.302 & 0.099 \\
 & Noise 40\% & 0.068 & 0.298 & 0.102 \\
  & Noise 80\% & 0.062 & 0.270 & 0.078 \\
 & Noise 100\% & 0.055 & 0.247 & 0.064 \\
\hline
 \multirow{6}{*}{Claude-4-Sonnet} 
& w/o RAG & 0.038 & 0.187 & 0.064 \\
 & Noise 0\% & 0.064 & 0.296 & 0.101 \\
 & Noise 20\% & 0.057 & 0.257 & 0.090 \\
 & Noise 40\% & 0.051 & 0.227 & 0.074 \\
  & Noise 80\% & 0.038 & 0.177 & 0.073 \\
 & Noise 100\% & 0.031 & 0.151 &  0.059 \\
\hline
\multirow{6}{*}{Gemini-2.5-Pro}
& w/o RAG & 0.055 & 0.272 & 0.092 \\
& Noise 0\% & 0.076 & 0.346 & 0.170 \\
 & Noise 20\% & 0.072 & 0.323 & 0.123 \\
 & Noise 40\% & 0.067 & 0.304 & 0.119 \\
  & Noise 80\% & 0.059 & 0.280 & 0.096 \\
 & Noise 100\% & 0.056 & 0.270 & 0.090 \\
\hline
\multirow{6}{*}{CitationGenerator-30B}
& w/o RAG & 0.043 & 0.225 & 0.165 \\
& Noise 0\% & 0.076 & 0.367 & 0.303 \\
& Noise 20\% & 0.056 & 0.298 & 0.242 \\
& Noise 40\% & 0.052 & 0.262 & 0.210 \\
& Noise 80\% & 0.047 & 0.240 & 0.175 \\
& Noise 100\% & 0.045 & 0.230 & 0.168 \\
\bottomrule
\end{tabular}
\end{adjustbox}
\label{tab:Robustness}
\end{table}

Table~\ref{tab:Robustness} evaluates model robustness against retrieval noise by introducing varying proportions of irrelevant papers into retrieved contexts. Closed-source models including GPT-5, Claude-4-Sonnet, and Gemini-2.5-Pro demonstrate moderate robustness, maintaining 80-90\% of their clean performance at 20\% noise levels but declining to 75-85\% at higher noise ratios. 
CitationGenerator-30B demonstrates superior robustness due to its fine-tuning, retaining approximately 80\% of clean performance even at 40\% noise and maintaining reasonable effectiveness at 80\% noise levels where other models collapse. 

\subsection{Use Case Analysis}
\label{app:usecase}

\begin{figure*}[!t]
    \centering
    \includegraphics[width=1.0\linewidth]{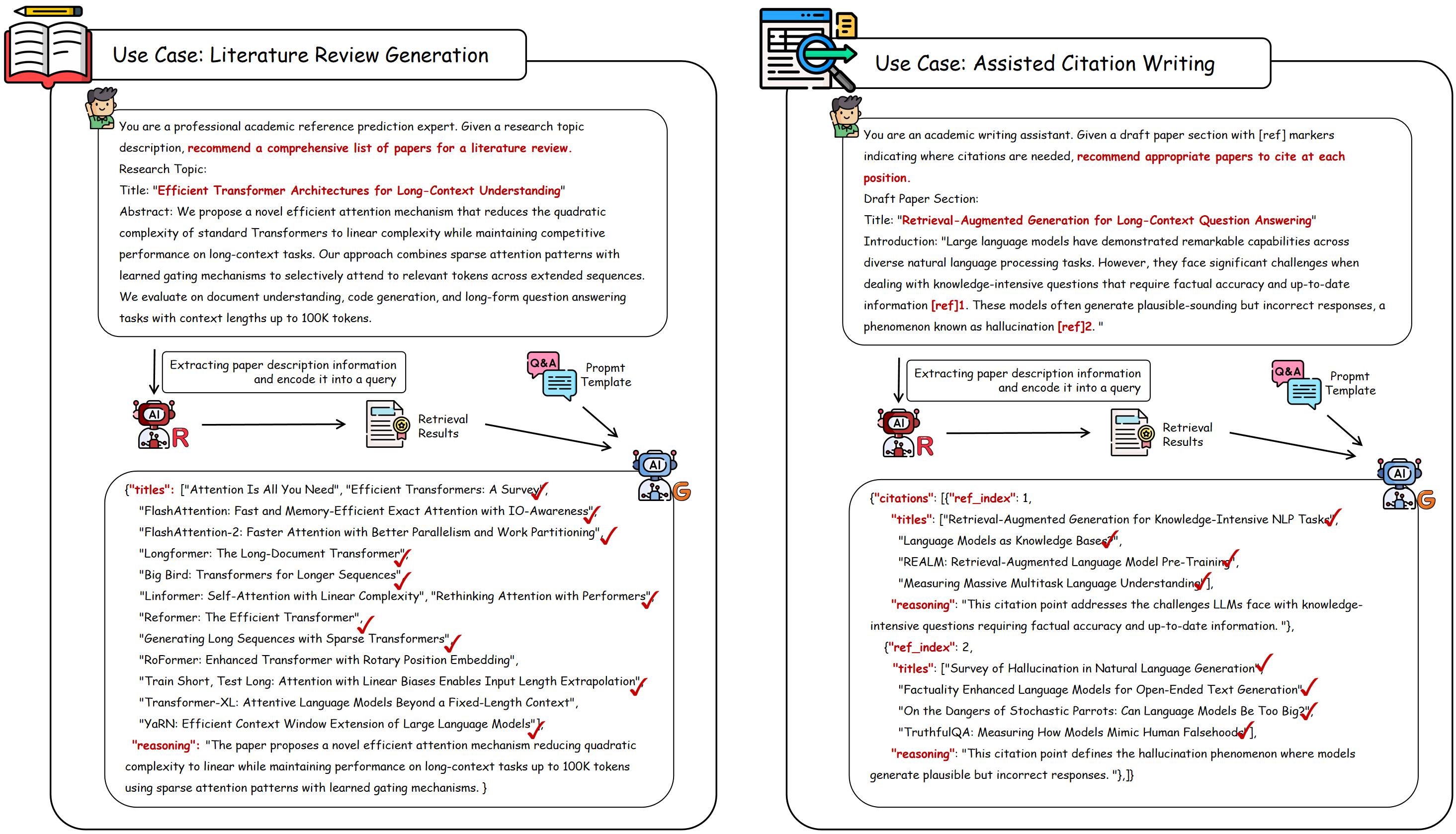}
    \vspace{-7mm}
    \caption{From Benchmark to Practice: Illustrative Use Cases of RAG-Enhanced Citation Prediction in Academic Research Workflows.}
    \vspace{-1mm}
\label{fig:usecase1}
\end{figure*}

Figure~\ref{fig:usecase1} demonstrates two representative applications corresponding to our benchmark tasks. The literature review generation use case aligns with Task 1's coarse-grained objective, where the system retrieves a comprehensive list of papers for a research topic on efficient Transformer architectures, with reasoning explaining each paper's relevance. The assisted citation writing use case corresponds to Task 2's fine-grained objective, where the system recommends specific papers for individual reference placeholders based on surrounding textual context, successfully identifying relevant works addressing hallucination and factuality challenges. These scenarios validate the practical utility of both task formulations for real-world academic workflows.

\subsection{Difficult Case Analysis}
\label{app:diff}

\begin{figure*}[!t]
    \centering
\vspace{-2mm}
    \includegraphics[width=1.0\linewidth]{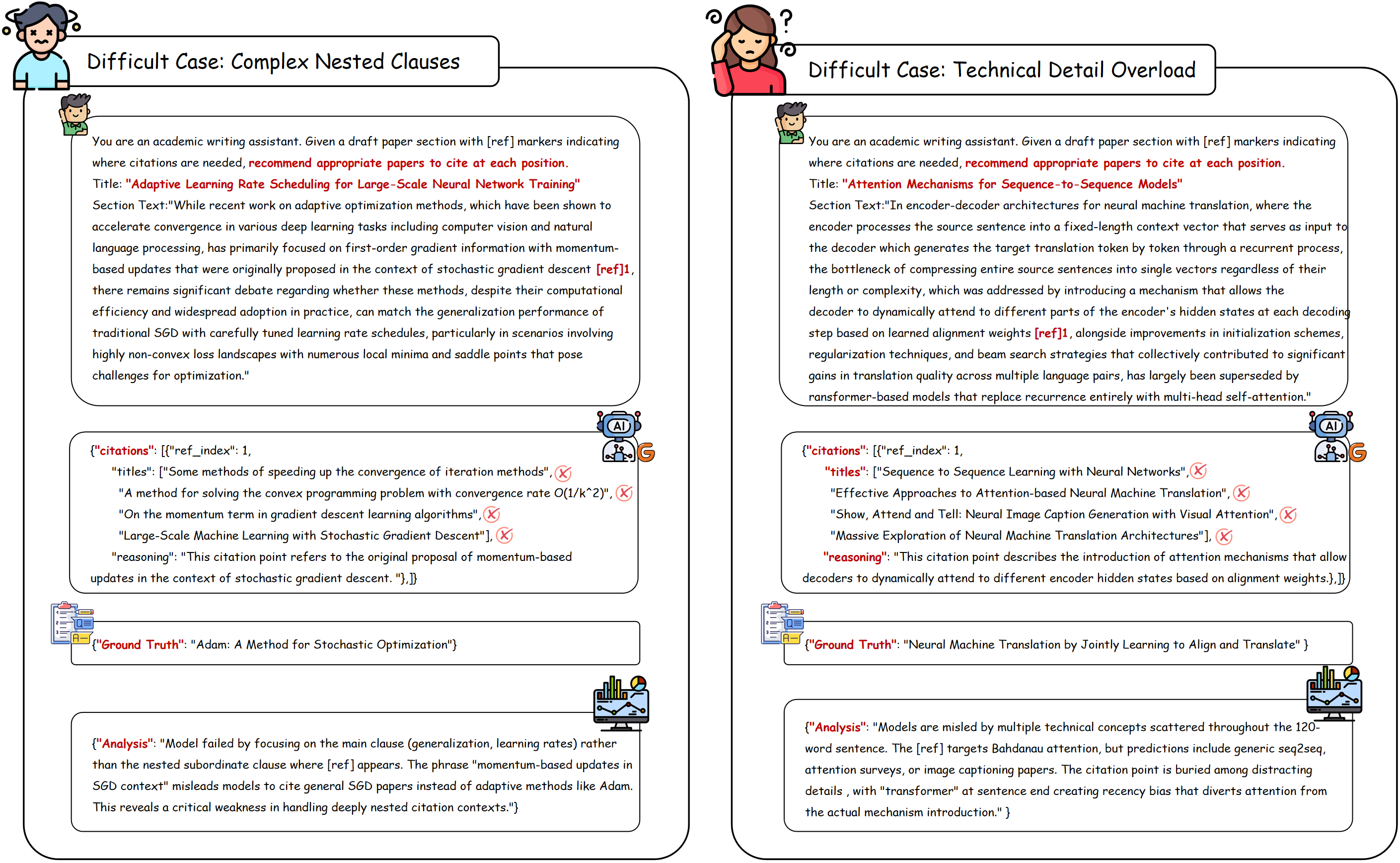}
    \vspace{-7mm}
    \caption{Difficult Case Analysis in Practice: Syntactic Complexity and Technical Overload as Critical Challenge Patterns.}
    \vspace{-4mm}
\label{fig:usecase2}
\end{figure*}

Figure~\ref{fig:usecase2} illustrates two challenging scenarios that expose current system limitations. The left panel shows a complex nested clause case where the citation appears within a deeply embedded subordinate clause, causing models to focus on the main clause about generalization rather than the nested context discussing momentum-based updates, resulting in incorrect predictions. The right panel demonstrates technical detail overload, where multiple scattered concepts in a lengthy sentence distract models from identifying the correct transformer mechanism reference, leading to recency bias toward superficial keyword matches. These failure cases highlight critical challenges in handling syntactically complex contexts and filtering relevant information from dense technical descriptions.

\clearpage
\newpage

\end{document}